\documentstyle[12pt]{article}

\makeatletter
\def\section{\@startsection {section}{1}{\z@}{-3.5ex plus -1ex minus
 -.2ex}{2.3ex plus .2ex}{\large\bf}}
\def\subsection{\@startsection{subsection}{2}{\z@}{-3.25ex plus -1ex minus
 -.2ex}{1.5ex plus .2ex}{\normalsize\bf}}
\makeatother

\makeatletter
\@addtoreset{equation}{section}

\makeatother

\newcommand{\nc}{\newcommand}
\newcommand{\rnc}{\renewcommand}

\nc{\be}{\begin{equation}}
\nc{\ee}{\end{equation}}
\nc{\bea}{\begin{eqnarray}}
\nc{\eea}{\end{eqnarray}}

\rnc{\a}{\alpha}
\rnc{\b}{\beta}
\rnc{\gg}{\gamma}
\rnc{\d}{\delta}
\nc{\e}{\eta}
\nc{\eb}{\bar{\eta}}
\nc{\ep}{\epsilon}
\nc{\f}{\phi}
\nc{\fb}{\bar{\phi}}
\nc{\vf}{\varphi}
\nc{\p}{\psi}
\rnc{\pb}{\bar{\psi}}
\rnc{\c}{\chi}
\nc{\cb}{\bar{\c}}
\nc{\la}{\lambda}
\nc{\m}{\mu}
\nc{\n}{\nu}
\rnc{\o}{\omega}
\nc{\Om}{\Omega}
\rnc{\t}{\theta}
\nc{\eps}{\epsilon}

\rnc{\S}{\Sigma}
\nc{\Sa}{\S\times\{0\}}
\nc{\Sb}{\S\times\{1\}}
\nc{\SI}{\S\times I}
\nc{\SS}{\S\times S^{1}}
\nc{\M}{{\cal M}}

\nc{\trac}[2]{{\textstyle\frac{#1}{#2}}}

\nc{\ex}[1]{{\rm e}^{\,\textstyle#1}}

\nc{\mat}[4]{\left(\begin{array}{cc}#1&#2\\#3&#4\end{array}\right)}

\nc{\som}[9]{\left(\begin{array}{ccc}#1&#2&#3\\#4&#5&#6\\#7&#8&#9%
\end{array}\right)}

\def\Tr{\mathop{\rm Tr}\nolimits}
\nc{\ra}{\rightarrow}
\nc{\Ra}{\Rightarrow}
\nc{\LRa}{\Leftrightarrow}
\nc{\ot}{\otimes}
\rnc{\ss}{\subset}
\nc{\nul}{\noindent\underline}

\nc{\subs}[1]{{\vspace*{0.5cm}}%
{\noindent\underline{#1}}{\addcontentsline{toc}{subsection}{#1}}%
{\vspace*{0.3cm}}}

\rnc{\lg}{{\bf g}}
\nc{\lt}{{\bf t}}
\nc{\lk}{{\lg/\lt}}
\nc{\lh}{{\bf h}}
\nc{\bft}{{\bf t}}
\nc{\bfk}{{\bf k}}
\nc{\bfg}{{\bf g}}
\nc{\del}{\partial}
\nc{\dz}{\del_{z}}
\nc{\dzb}{\del_{\bar{z}}}
\nc{\zb}{\bar{z}}
\nc{\az}{A_{z}}
\nc{\azb}{A_{\bar{z}}}
\nc{\bz}{B_{z}}
\nc{\bzb}{B_{\bar{z}}}
\nc{\ba}{{\bf A}}
\nc{\bb}{{\bf B}}
\nc{\ah}{A^{h}}
\nc{\aht}{(\ah)^{\lt}}
\nc{\ahk}{(\ah)^{\lg/\lt}}
\nc{\g}{g^{-1}}
\nc{\dw}{\Delta_{W}}
\nc{\Det}{{\rm Det}\,}
\nc{\Ad}{{\rm Ad}}
\nc{\bG}{{\bf G}}
\nc{\bT}{{\bf T}}
\nc{\bK}{{\bf H}}
\nc{\bH}{{\bf H}}
\nc{\bP}{{\bf P}}
\nc{\gt}{\bG/\bT}
\nc{\map}[2]{{\rm Map}(#1,#2)}
\nc{\mg}{\map{M}{\bG}}
\nc{\mgr}{\map{M}{\bG_{r}}}
\nc{\mt}{\map{M}{\bT}}
\nc{\mtr}{\map{M}{\bT_{r}}}
\nc{\sg}{\map{\S}{\bG}}
\nc{\sgr}{\map{\Sigma}{\bG_{r}}}
\nc{\st}{\map{\Sigma}{\bT}}
\nc{\str}{\map{\Sigma}{\bT_{r}}}
\nc{\mgt}{\map{M}{\gt}}
\nc{\sgt}{\map{\Sigma}{\gt}}
\nc{\ug}{\map{U}{\bG}}
\nc{\ugr}{\map{U}{\bG_{r}}}
\nc{\ut}{\map{U}{\bT}}
\nc{\utr}{\map{U}{\bT_{r}}}
\nc{\uag}{\map{\ua}{\bG}}
\nc{\uagr}{\map{\ua}{\bG_{r}}}
\nc{\uat}{\map{\ua}{\bT}}
\nc{\uatr}{\map{\ua}{\bT_{r}}}
\nc{\uabg}{\map{\uab}{\bG}}
\nc{\uabgr}{\map{\uab}{\bG_{r}}}
\nc{\uabt}{\map{\uab}{\bT}}
\nc{\uabtr}{\map{\uab}{\bT_{r}}}
\nc{\unA}{\underline{A}}
\nc{\C}{{\cal A}/{\cal G}}
\nc{\cA}{{\cal A}}
\nc{\ag}{\cA(g)}
\nc{\ng}{{\cal N}(g)}
\nc{\xa}{X_{\cA}}
\nc{\xag}{\xa(g)}
\nc{\dx}{\dot{x}}
\rnc{\O}[2]{\Omega^{#1}({#2},\lg)}
\nc{\wif}{Weyl integral formula}
\nc{\CS}{Chern-Simons theory}
\nc{\tg}{\tilde{g}}
\nc{\tti}{\tilde{t}}
\nc{\th}{\tilde{h}}

\nc{\ga}{g_{\a}}
\nc{\gb}{g_{\b}}
\nc{\gc}{g_{\gg}}
\nc{\ha}{h_{\a}}
\nc{\hb}{h_{\b}}
\nc{\hc}{h_{\gg}}
\nc{\ka}{k_{\a}}
\nc{\kb}{k_{\b}}
\nc{\kc}{k_{\gg}}
\nc{\ta}{t_{\a}}
\nc{\tb}{t_{\b}}
\nc{\tc}{t_{\gg}}
\nc{\gab}{g_{\a\b}}
\nc{\gac}{g_{\a\gg}}
\nc{\gbc}{g_{\b\gg}}
\nc{\hab}{h_{\a\b}}
\nc{\hac}{h_{\a\gg}}
\nc{\hbc}{h_{\b\gg}}
\nc{\kab}{k_{\a\b}}
\nc{\kac}{k_{\a\gg}}
\nc{\kbc}{k_{\b\gg}}
\nc{\tab}{t_{\a\b}}
\nc{\tac}{t_{\a\gg}}
\nc{\tbc}{t_{\b\gg}}
\nc{\ua}{U_{\a}}
\nc{\ub}{U_{\b}}
\nc{\uc}{U_{\gg}}
\nc{\uab}{U_{\a}\cap U_{\b}}
\nc{\uac}{U_{\a}\cap U_{\gg}}
\nc{\ubc}{U_{\b}\cap U_{\gg}}

\def\sAA{{\rm A\kern-0.85em A}} 
\def\tAA{{\mathchoice
  {\sAA}
  {\sAA}
  {\rm A\kern-0.60em A}
  {\rm A\kern-0.50em A} }}
\def\sBB{{\rm I\kern-.17em{}B}}
\def\BB{{\mathchoice
  {\sBB}
  {\sBB}
  {\rm I\kern-.13em{}B}
  {\rm I\kern-.13em{}B} }}
\def\sCC{{\kern 0.27em\vrule height1.45ex width0.03em depth0em
          \kern-0.30em\rm C}}
\def\CC{{\mathchoice
  {\sCC}
  {\sCC}
  {\kern 0.225em \vrule height1.05ex width0.025em depth0em \kern-0.25em \rm C}
  {\kern 0.180em \vrule height0.78ex width0.02em depth0em \kern-0.2em \rm C}
        }}
\def\tCC{{\ooalign{C\crcr\kern0.27em\vrule height1.45ex width0.03em
depth0em\crcr}}}
\def\sDD{{\rm I\kern-.16em{}D}}
\def\DD{{\mathchoice
  {\sDD}
  {\sDD}
  {\rm I\kern-.13em{}D}
  {\rm I\kern-.13em{}D} }}
\def\sEE{{\rm I\kern-.17em{}E}}
\def\EE{{\mathchoice
  {\sEE}
  {\sEE}
  {\rm I\kern-.13em{}E}
  {\rm I\kern-.13em{}E} }}
\def\sFF{{\rm I\kern-.16em{}F}}
\def\FF{{\mathchoice
  {\sFF}
  {\sFF}
  {\rm I\kern-.13em{}F}
  {\rm I\kern-.13em{}F} }}
\def\sGG{{\kern 0.27em \vrule height1.45ex width0.03em depth0em
          \kern-0.30em \rm G}}
\def\GG{{\mathchoice
  {\sGG}
  {\sGG}
  {\kern 0.225em \vrule height1.05ex width0.025em depth0em \kern-0.25em \rm G}
  {\kern 0.180em \vrule height0.78ex width0.020em depth0em \kern-0.20em \rm G}
        }}
\def\sHH{{\rm I\kern-.16em{}H}}
\def\HH{{\mathchoice
  {\sHH}
  {\sHH}
  {\rm I\kern-.13em{}H}
  {\rm I\kern-.13em{}H} }}
\def\sII{{\rm I\kern-.16em{}I}}
\def\II{{\mathchoice
  {\sII}
  {\sII}
  {\rm I\kern-.12em{}I}
  {\rm I\kern-.10em{}I} }}
\def\sJJ{{\kern0.17em\vrule height1.5ex width 0.03em depth0em
          \kern-.20em\rm J}}
\def\JJ{{\mathchoice
  {\sJJ}
  {\sJJ}
  {\kern0.150em\vrule height1.05ex width 0.025em depth0em\kern-.175em\rm J}
  {\kern0.135em\vrule height0.78ex width 0.020em depth0em\kern-.155em\rm J} }}
\def\sKK{{\rm I\kern-.16em{}K}}
\def\KK{{\mathchoice
  {\sKK}
  {\sKK}
  {\rm I\kern-.12em{}K}
  {\rm I\kern-.10em{}K} }}
\def\sLL{{\rm I\kern-.16em{}L}}
\def\LL{{\mathchoice
  {\sLL}
  {\sLL}
  {\rm I\kern-.12em{}L}
  {\rm I\kern-.10em{}L} }}
\def\sMM{{\rm I\kern-.16em{}M}}
\def\MM{{\mathchoice
  {\sMM}
  {\sMM}
  {\rm I\kern-.12em{}M}
  {\rm I\kern-.10em{}M} }}
\def\sNN{{\rm I\kern-.16em{}N}}
\def\NN{{\mathchoice
  {\sNN}
  {\sNN}
  {\rm I\kern-.12em{}N}
  {\rm I\kern-.10em{}N} }}
\def\sOO{{\kern 0.27em \vrule height1.50ex width0.03em depth0em
					\kern-0.30em \rm O}}
\def\OO{{\mathchoice
  {\sOO}
  {\sOO}
  {\kern 0.225em \vrule height1.05ex width0.025em depth0em \kern-0.25em \rm O}
  {\kern 0.180em \vrule height0.78ex width0.020em depth0em \kern-0.20em \rm O}
        }}
\def\sPP{{\rm I\kern-.16em{}P}}
\def\PP{{\mathchoice
  {\sPP}
  {\sPP}
  {\rm I\kern-.12em{}P}
  {\rm I\kern-.10em{}P} }}
\def\sQQ{{\kern 0.27em \vrule height1.45ex width0.03em depth0em
          \kern-0.30em \rm Q}}
\def\QQ{{\mathchoice
	{\sQQ}
	{\sQQ}
  {\kern 0.225em \vrule height1.05ex width0.025em depth0em \kern-0.25em \rm Q}
  {\kern 0.180em \vrule height0.78ex width0.020em depth0em \kern-0.20em \rm Q}
        }}
\def\sRR{{\rm I\kern-0.16em{}R}}
\def\RR{{\mathchoice
  {\sRR}
  {\sRR}
  {\rm I\kern-0.12em{}R}
  {\rm I\kern-0.10em{}R} }}
\def\sSS{{\rm S\kern-.45em{}S}}
\def\sTT{{\rm T\kern-.60em{}T}}
\def\TT{{\mathchoice
  {\sTT}
  {\sTT}
  {\rm T\kern-.45em{}T}
  {\rm T\kern-.38em{}T} }}
\def\sUU{{\rm U\kern-.60em{}U}}
\def\UU{{\mathchoice
  {\sUU}
  {\sUU}
  {\rm U\kern-.46em{}U}
  {\rm U\kern-.38em{}U} }}
\def\sVV{{\rm V\kern-.62em{}V}}
\def\VV{{\mathchoice
  {\sVV}
  {\sVV}
  {\rm V\kern-.46em{}V}
  {\rm V\kern-.38em{}V} }}
\def\sWW{{\rm W\kern-.92em{}W}}
\def\WW{{\mathchoice
  {\sWW}
  {\sWW}
  {\rm W\kern-.80em{}W}
  {\rm W\kern-.67em{}W} }}
\def\sXX{{\rm X\kern-.58em{}X}}
\def\XX{{\mathchoice
  {\sXX}
  {\sXX}
  {\rm X\kern-.45em{}X}
  {\rm X\kern-.38em{}X} }}
\def\sYY{{\rm Y\kern-.58em{}Y}}
\def\YY{{\mathchoice
  {\sYY}
  {\sYY}
  {\rm Y\kern-.45em{}Y}
  {\rm Y\kern-.40em{}Y} }}
\def\sZZ{{\rm Z\kern-0.32em{}Z}}
\def\ZZ{{\mathchoice
  {\sZZ}
  {\sZZ}
  {\rm Z\kern-0.30em{}Z}
  {\rm Z\kern-0.25em{}Z} }}

\begin{document}
\global\parskip=4pt


\begin{titlepage}
\newlength{\titlehead}
\settowidth{\titlehead}{ENSLAPP-L-xxx/xx}
\begin{flushright}
\parbox{\titlehead}{
\begin{flushleft}
IC/94/108\\
ENSLAPP-L-469/94\\
June 1994\\
\end{flushleft}
}
\end{flushright}
\begin{center}
\vskip .25in
{\LARGE\bf Equivariant K\"ahler Geometry and} \\
\vskip .25in
{\LARGE\bf Localization in the $\bf G/G$ Model}\\
\vskip .25in
{\bf Matthias Blau}\footnote{e-mail: blau@ictp.trieste.it}
and {\bf George Thompson}\footnote{e-mail: thompson@ictp.trieste.it}\\
\vskip .10in
ICTP\\
P.O. Box 586\\
I-34014 Trieste\\
Italy
\end{center}
\vskip .15in
\begin{abstract}
We analyze in detail the equivariant supersymmetry of the $G/G$ model.
In spite of the fact that this supersymmetry does not model the infinitesimal
action of the group of gauge transformations, localization can be established
by standard arguments. The theory localizes onto reducible connections and a
careful evaluation of the fixed point contributions leads to an alternative
derivation of the Verlinde formula for the $G_{k}$ WZW model.

We show that the supersymmetry of the $G/G$ model can be regarded as an
infinite dimensional realization of Bismut's theory of equivariant Bott-Chern
currents on K\"ahler manifolds, thus providing a convenient cohomological
setting for understanding the Verlinde formula.

We also show that the supersymmetry is related to a non-linear generalization
($q$-deformation) of the ordinary moment map of symplectic geometry in which a
representation of the Lie algebra of a group $G$ is replaced by a
representation
of its group algebra with commutator $[g,h] = gh-hg$. In the large $k$ limit it
reduces to the ordinary moment map of two-dimensional gauge theories.
\end{abstract}
\end{titlepage}


\begin{small}
\tableofcontents
\end{small}

\setcounter{footnote}{0}

\section{Introduction}

In \cite{btver} we showed how to obtain the Verlinde formula \cite{ver} for the
dimension of the space of conformal blocks of the $\bG_{k}$ Wess-Zumino-Witten
model by explicit evaluation of the partition function of the $\bG_{k}/\bG_{k}$
model using Abelianization, i.e.~a functional integral version of the Weyl
integral formula for compact Lie groups. This Abelianization could
alternatively be regarded as a localization of the path integral, although
the suspersymmetric structure of equivariant cohomolgy usually responsible for
such a localization was not manifest in \cite{btver}.

On the other hand, in
\cite{ger,btlec} it was pointed out that the $G/G$ model has a supersymmetric
extension analogous to that (and useful for the cohomological interpretation)
of Yang-Mills theory \cite{ew2d,btlec}. While the supersymmetry $\d$ is
somewhat
unusual in that it does not square to infinitesimal gauge transformations and
hence does not model the action of the gauge group on the space of fields,
we want to emphasize that in
principle the localization theorems also apply in this situation and can be
used to evaluate the partition function. The reason for this is that if the
action contains a term of the form $\d f$ with $\d^{2}f=0$ then the
functional integral is (formally) independent of the coefficient of this
term and hence the addition of such terms to the action
can be used to localize the integral without changing its value. If $\d$
happens to square to infinitesimal gauge transformations, then such functionals
$f$ are easy to find (the requirement $\d^{2}f=0$ amounting to the gauge
invariance of $f$), while in general this may be more difficult.
In the $G/G$ model, though, one does not have to look very far as one
is in the fortunate situation where the classical action itself already
contains a term of this type whose coefficient can then be varied to establish
localization.

That the Verlinde formula should
have an interpretation as a fixed point formula had been suggested long ago
\cite{verver} on the basis of its algebraic structure, and here we find a
manifestation of this at the path integral level. The link with the method used
in \cite{btver} is provided by the observation that localization with respect
to this supersymmetry essentially abelianizes the theory in the sense that it
localizes to reducible connections. The detailed path integral
argument for localization turns out to be slightly more complicated than
a simple stationary phase approximation argument would suggest, which is why
we present it in some detail here both for the $G/G$ model and, in an appendix,
for Yang-Mills theory. But although there are also some subtleties
related e.g.~to obstructions to the global diagonalizability of group valued
maps, explored from a mathematical point of view in \cite{btdia}, that part of
the story is nevertheless quite straightforward (a rough, although not quite
correct and to the point, sketch of the argument having already been given in
\cite{ger}) and, all by itself, not terribly enlightning.

What we want to mainly draw attention to in this paper is that this
supersymmetry actually encodes a much richer and more interesting structure,
both from the complex holomorphic and the equivariant symplectic point of view,
than would be required for localization alone. First of all, although the
supersymmetry is not nilpotent, satisfying e.g.
\[\d^{2} A_{z} = A_{z}^{g}-A_{z}\;\;,\;\;\;\;\;\;
  \d^{2}A_{\zb}=A_{\zb}-A_{\zb}^{g^{-1}}\;\;,\]
it can be split into a sum of two nilpotent operators $Q$ and $\bar{Q}$,
\[\delta = Q + \bar{Q}\;\;,\;\;\;\;\;\;Q^{2}=\bar{Q}^{2} = 0\;\;.\]
These operators can be regarded as equivariant Dolbeault operators with
respect to a ($g$-dependent) holomorphic Killing vector field $X$ on the
space $\cA$ of gauge fields,
\[Q = \del_{\cA} + i(X^{(0,1)})\;\;,\;\;\;\;\;\;
\bar{Q}=\bar{\del}_{\cA} + i(X^{(1,0)})\;\;.\]
Although the $G/G$ action itself is not manifestly topological, it splits
naturally into a $Q\bar{Q}$-exact part and a cohomologically non-trivial term,
the latter being the manifestly topological gauged Wess-Zumino term.
Formally, the theory should then be independent of the coefficient of the
former and in this way we recover the one-parameter family of deformations
of the $G/G$ model discussed by Witten \cite{ew}. The supersymmetric extension
we consider here automatically keeps track of the required quantum corrections
to ensure the constancy of this one-parameter family of theories also at the
quantum level.

Modulo the usual one-loop determinant, the calculation of the partition
function then reduces essentially
to the evaluation of the gauged Wess-Zumino term
$\Gamma(g,A)$ for reducible configurations $A^{g}=A$. For fixed $g$,
$\Gamma(g,A)$ turns out to be independent of $A$ and related to the generalized
winding numbers introduced in \cite{btdia}. This relates
the Verlinde formula to Chern classes of torus bundles and provides another
manifestation of the Abelianization inherent in the Verlinde formula.

Interestingly, the gauge field functional integral of the supersymmetric
extension of the $G/G$ model is precisely of the form of the integrals
studied by Bismut \cite{bismut} in his investigations of the relations
among complex equivariant cohomology, Ray-Singer torsion, anomaly formulae
for Quillen metrics and equivariant Bott-Chern currents. Here we make this
analogy precise in the belief that it provides a convenient, and from other
points of view not completely obvious, cohomological setting for understanding
the Verlinde formula. In particular, it identifies the above winding numbers
as equivariant cohomology classes on the space of connections.

What is still missing to complete the picture is a direct demonstration that
the $G/G$ functional integral represents the Riemann-Roch integral over the
moduli space of flat connections for the dimension of the space of conformal
blocks (or holomorphic sections of some power of the determinant line bundle).
In particular, both in the approach pursued in this paper and in the one
based on Abelianization, localization onto flat connections is conspicuously
absent at every stage of the calculation. One possibility would be to try to
find a cohomological topological field theory which has the same relation to
the $G/G$ model that 2d Donaldson theory has to BF (topological Yang-Mills)
theory \cite{ew2d}. Finding such an alternative localization should also
provide one directly with a finite dimensional integral which yields the
Verlinde formula via some fixed point theorem or localization formula,
but our attempts in this direction have as of yet been unsuccessful.

All this is more or less analogous to the situation in mathematics where such
a direct proof of the Verlinde formula is also still missing (see \cite{beau}
for an up-to-date account of the mathematical status of the Verlinde formula),
while Szenes \cite{szenes} has indicated how it would follow from a
proof \cite{jk} of the Witten conjectures \cite{ew2d} on the cohomology
of the moduli space of flat connections.

The somewhat unusual supersymmetry of the $G/G$ model also leads to a
modification of the underlying symplectic geometry. The $g$-dependent vector
fields $X=X(g)$ on $\cA$ satisfy the algebra
\[ [X(g),X(h)] = X(gh)-X(hg)\;\;.\]
In particular, therefore, they do not provide a representation of the
Lie algebra of the gauge group on $\cA$, but rather of its group algebra
equipped with the Lie bracket $[g,h]=gh-hg$. These vector fields are
Hamiltonian, the Hamiltonian (or moment map) being the $G/G$ action $S(g,A)$
itself. This moment map is equivariant in the sense that the Lie bracket
relation among the vector fields can be lifted to the Poisson algebra of
function(al)s on $\cA$ - in fact, equivariance turns out to be equivalent
to the Polyakov-Wiegmann identity and this fixes the $A$-independent part
of the Hamiltonian $S(g,A)$ up to a natural ambiguity. Hence the above
translates into the Poisson bracket relation
\[\{S(g,A),S(h,A)\} = S(gh,A)-S(hg,A)\]
for the $G/G$ action. This moment map with its generalized equivariance,
the Lie algebra having been replaced by the group algebra, is a deformation
of the ordinary equivariant moment map of two-dimensional gauge theories in
the sense that it reduces to it in the $k\ra\infty$ limit where the
$G/G$ action at level $k$ becomes the BF action. The latter is
nothing other than the generator of ordinary
gauge transformations on the space of gauge fields.

This paper is organized as follows. In section 2 we discuss various
aspects of the supersymmetric extension of the $G/G$ model and its
one-parameter family of deformations. The follownig two sections can
then be read fairly independently of each other.
In section 3 we first describe the
relevant aspects of Bismut's theory of Bott-Chern currents as well as
the localization theorem for their integrals. We then investigate in some
detail the path integral argument leading to the localization of the
partition function of the $G/G$ model to the `classical' set of
reducible configurations. The corresponding argument for Yang-Mills theory
as well as some alternative strategies are discussed in Appendix A.
At this point the intermediate expression for the partition
function one obtains is identical to that arrived at in \cite{btver} upon
Abelianization and we therefore only sketch briefly how everything can be
put together to obtain the Verlinde formula, referring to \cite{btver,btlec}
for details. We begin section 4 with a brief review of ordinary Hamiltonian
group actions, show that the $G/G$ action can be interpreted as a
moment map satisfying the above generalized equivariance condition,
discuss the $k\ra\infty$ limit and finally extract from the preceding
discussion the basic structure of generalized Hamiltonian group actions
and the relation with the standard theory.

\section{The Supersymmetry of the $\bf G/G$ Model}

We begin with a brief review of those aspects of the $G/G$ model which
are of relevance to us. The action of the $G/G$ model at level $k\in\ZZ$
is
\bea
kS_{G/G}(g,A)&=& kS_{G}(g,A) -ik\Gamma(g,A)\label{diff}\;\;,\\
S_{G}(g,A) &=&-\trac{1}{8\pi}\int_{\S} \g d_{A}g *\g d_{A}g
\;\;,\label{diff1}\\
\Gamma(g,A)&=& \trac{1}{12\pi}\int_{N} (\g dg)^{3}-\trac{1}{4\pi}\int_{\S}
\left(A\,dg\,\g + A A^{g}\right)\;\;.\label{diff2}
\eea
Here $g\in{\cal G}=\sg$ is a (smooth) group valued field on a two-dimensional
closed surface $\S$ (with an extension to
a bounding three manifold $N$ in the Wess-Zumino term
$\Gamma(g) = \Gamma(g,A=0)$). Not aiming for maximal generality, we will
assume that $\bG$ is simply connected. $A$ is a gauge
field for the diagonal $\bG$ subgroup of the $\bG_{L}\times\bG_{R}$ symmetry
of the ungauged WZW action $S_{G}(g)=S_{G/G}(g,A=0)$. The covariant
derivative is $d_{A}g = dg + [A,g]$, $A^{g} = \g A g + \g dg$ is the gauge
transform of $A$, and $*$ is the Hodge duality operator with respect
to some metric on $\S$. Acting on one-forms, $*$ is conformally invariant
so that the action only depends on a complex structure on $\S$.
In the above formulae and in the following, integrals of Lie algebra valued
forms are understood to include a trace.
We will occasionally find it convenient to split this action into its
$A$-independent and $A$-dependent part as
$S_{G/G}(g,A)=S_{G}(g)+S_{/G}(g,A)$.

\subs{Symmetries and Equations of Motion of the $G/G$ Model}

We now list some properties of the $G/G$ model we
will make use of below. First of all, by construction, the action is
invariant under the local gauge transformations
\be
g\ra g^{h}\equiv h^{-1}g h \;\;,\;\;\;\;\;\;A\ra A^{h}\equiv
h^{-1}Ah + h^{-1}dh \;\;.\label{wz3}
\ee
The variation of the action $S_{G/G}$ with respect to the gauge fields is
\be
\d S_{G/G}(g,A) = \trac{1}{2\pi}\int_{\S}  (J_{z} \d\azb - J_{\zb}\d\az)\;\;,
\label{deltaa}
\ee
where $J_{z}$ and $J_{\zb}$ are the covariantized versions
\be
J_{z}= \g D_{z} g = A_{z}^{g}-A_{z}\;\;,\;\;\;\;\;\;
J_{\zb} = D_{\zb}g\,\g = A_{\zb}-A_{\zb}^{\g}\;\;,
\ee
of the currents $j_{z}=\g\dz g$ and $j_{\zb}=\dzb g\,\g$ generating the
Kac-Moody symmetry of the WZW model $S_{G}(g)$.
Since they are gauge currents, they are set to zero by the equations of
motion of the gauge fields. An equivalent way of expressing the
vanishing of the current $J=J_{z}dz + J_{\zb}d\zb$ is
\be
J_{z}=J_{\zb}=0\Leftrightarrow d_{A}g = 0 \Leftrightarrow A^{g}=A\;\;.
\ee
The remaining equation of motion can then be cast into the form $F_{A}=0$
so that classical configurations are gauge equivalence classes of pairs
$(A,g)$ where $A$ is flat and $g$ is a symmetry of $A$.
This is very reminiscent of the phase space of Chern-Simons theory on
a three-manifold of the form $\S\times\RR$ and even more of that
of two-dimensional non-Abelian BF theory (see \cite{btlec} for a detailed
comparison of these two theories). This already suggests that the
$G/G$ model is a topological field theory and this can indeed be established,
either by showing directly that the variation of the partition function with
respect to the metric is zero \cite{ewwzw} or by referring to the equivalence
of the $G/G$ model with Chern-Simons theory on $\SS$ established in
\cite{btver}.  Yet another argument will follow from our considerations below,
concerning the relation between the $G/G$ model and the manifestly topological
theory with action the gauged WZ term $\Gamma(g,A)$.

For later use, we note here a cocycle identity
satisfied by the action of the $G/G$ model.
It is a generalization of the Polyakov-Wiegmann identity
\be
S_{G}(gh)=S_{G}(g)+S_{G}(h)-\trac{1}{2\pi}\int_{\S} j_{z}(g)j_{\zb}(h)
\label{pw1}
\ee
for the WZW action and reads
\be
S_{G/G}(gh,A) = S_{G/G}(g,A) + S_{G/G}(h,A)
                -\trac{1}{2\pi}\int_{\S}J_{z}(g)J_{\zb}(h)\label{pw2}\;\;.
\ee
This ends our review of the $G/G$ model and now we turn to those aspects
of the theory related to supersymmetry and the (equivariant) K\"ahler
geometry on the space of fields.

\subs{The Supersymmetric Extension of the $G/G$ Action}

In the case of BF theory and 2d Yang-Mills theory it was found \cite{ew2d}
that the geometric interpretation of the theory was greatly facilitated
by adding to the original bosonic action a term $\sim\int_{\S}\p_{z}\p_{\zb}$
quadratic in the Grassmann odd variables $\p$ and representing the
symplectic form $\sim\int_{\S}\d A\d A$ on the space $\cA$ of gauge fields
on $\S$. The resulting theory turned out to be supersymmetric and the
supersymmetry could be interpreted as a representation of equivariant
cohomology with respect to the infinitesimal action of the gauge group on
$\cA$. Something analogous is also possible (and turns out to be useful)
here, and we just want to mention in passing that a similar supersymmetry
can also be shown to exist in Chern-Simons theory on $\S\times S^{1}$.

As a consequence of (\ref{deltaa}) the combined action
\bea
S(g,A,\p) &=& S_{G/G}(g,A) -  \Omega(\p)\label{susact} \\
\Omega(\p) &=& \trac{1}{2\pi}\int_{\S} \p_{z}\p_{\zb}\label{omega}
\eea
is invariant under the (supersymmetry) transformations
\bea
&&\d\az = \p_{z}\;\;,\;\;\;\;\;\;\d\p_{z} = J_{z} \;\;,
\nonumber\\
&&\d\azb= \p_{\zb}\;\;,\;\;\;\;\;\;\d\p_{\zb} = J_{\zb}\;\;,
\label{ggsusy1}
\eea
supplemented by $\d g =0$. Note that this is a complex transformation,
$\p_{z}$ and $\p_{\zb}$ not transforming as complex conjugates of each
other. That such a transformation can nevertheless be a symmetry of the
action is due to the
fact that in Euclidean space the action of the (gauged) WZW model is
itself complex, the imaginary part being given by the (gauged) WZ term.

What is interesting about this supersymmetry is
that, unlike its Yang-Mills counterpart, it does not square to infinitesimal
gauge transformations but rather to `global' or `large' gauge transformations,
\bea
\d^{2}\az = \az^{g} - \az\;\;,&&\d^{2}\p_{z} = \p_{z}^{g}-\p_{z}\equiv
\g\p_{z} g - \p_{z} \;\;,\nonumber\\
\d^{2}\azb = \azb - \azb^{\g}\;\;,&&\d^{2}\p_{\zb} = \p_{\zb}-\p_{\zb}^{\g}
\equiv \p_{\zb} - g\p_{\zb} \g\;\;. \label{ggsusy2}
\eea
In particular, this implies that, in addition to (infinitesimal)
gauge invariance, the purely
bosonic action $S_{G/G}$ has another {\em infinitesimal} invariance $\Delta$
given by $\Delta = \d^{2}$,
\be
\Delta A = J \Rightarrow \Delta S_{G/G}(g,A) = 0\;\;.
\ee
It is, however, a rather trivial symmetry from another point of view as
it is simply proportional to the classical equations of motion and as such
a symmetry present in any action: if $S(\Phi^{k})$ is a functional of the
fields $\Phi^{k}$ and one defines a variation of $\Phi^{k}$ by
\be
\Delta\Phi^{k}= \epsilon^{kl}\frac{\d S}{\d\Phi^{l}} \label{Delta}
\ee
where $\epsilon^{kl}$ is antisymmetric, then the action is invariant,
\be
\Delta S = \epsilon^{kl}\frac{\d S}{\d \Phi^{k}}\frac{\d S}{\d\Phi^{l}}=0\;\;.
\ee
Actually, also the supersymmetry (\ref{ggsusy1}) itself can be regarded as
such a trivial symmetry of the extended action (\ref{susact}) as $\p_{z}$
acts as a source for $\p_{\zb}$ and vice-versa. From this point of view it
is perhaps even more surprising that this supersymmetry is nevertheless a
useful symmetry to consider. Partly this is due to the fact that, while
the bosonic symmetry is only an infinitesimal symmetry  (its exponentiated
version involving higher derivatives of the action), its fermionic version
is a full symmetry of the action as it stands.

In the case of Yang-Mills theory and BF theory, the two infinitesimal
symmetries, gauge transformations and $\Delta$, coincide whereas here
the supersymmetry does not model the standard equivariant cohomology on the
space of gauge fields. While this does not preclude localization (which can,
after all, be established for any Killing vector field on a symplectic
manifold \cite{bv}), it does lead to certain unusual features and some
care has to be exercised when adapting the usual arguments establishing
localization of the partition function to the present case. In particular,
it suggests that some global (or rather, as we will see, $q$-deformed)
counterpart of ordinary infinitesimal
equivariant cohomology could provide the right interpretational
framework for this model - an issue that appears to merit further
investigation. In section 4 we will explore some of the symplectic geometry
involved. In the following, however, we will focus on another geometrical
framework, related to the theory of equivariant Bismut-Bott-Chern currents
\cite{bismut}, a framework which seems to be particularly well adapted to
the study of the $G/G$ model and within which localization can be
established along similar lines as in the case of ordinary equivariant
cohomology.

\subs{The Supersymmetry and Holomorphic Killing Vector Fields on $\cA$}

We start by rewriting the
supersymmetry in a slightly more familiar form.
The supersymmetry operator $\d$ can be written as the sum
of two nilpotent Dolbeualt like operators $Q$ and $\bar{Q}$,
\be
\d = Q + \bar{Q}\;\;,\;\;\;\;\;\; Q^{2} = \bar{Q}^{2} = 0\;\;, \label{qq0}
\ee
where e.g.
\bea
QA_{z} = \p_{z}\;\;&,&\;\;\;\;\;\;Q \azb =0 \;\;, \nonumber\\
Q\p_{z} = 0 \;\;&,&\;\;\;\;\;\; Q\p_{\zb} = J_{\zb}\;\;.
\eea
The action $S_{G/G}$ is also seperately $Q$ and $\bar{Q}$ invariant.
We should perhaps properly write $\d=\d(g)$ and $Q=Q(g)$ and we will
occasionally
do this when we find it necessary to emphasize that there is not just one
but rather a whole $\cal G$'s worth of these derivations on $\cA$.

To gain some insight into the geometrical meaning of this supersymmetry,
we introduce the $g$-dependent vector field
\be
\xag = J_{z}(g)\frac{\d}{\d\az} + J_{\zb}(g)\frac{\d}{\d\azb} \equiv
       \xag^{(1,0)} + \xag^{(0,1)}             \label{xag}
\ee
on $\cA$ (actually a section of the complexified tangent bundle of $\cA$).
Note that the infinitesimal action of this vector field generates
a global chiral gauge transformation, in the sense that
e.g. $\xag\az=\az^{g}-\az$, so that the exponentiated action takes the
form of an iterated (chiral) gauge transformation if $\az$ and $\azb$
are treated as independent fields.

Denoting the exterior deriviative on $\cA$ by $d_{\cA}$ and
contraction by a vectorfield $Y$ by $i(Y)$ we can represent
the supersymmetry $\d$ as the equivariant exterior derivative on
$\cA$ with respect to $\xag$,
\be
\d(g)=d_{\cA} + i(\xag)\;\;,                              \label{dxag}
\ee
and $Q(g)$ and $\bar{Q}(g)$ by
\be
Q(g)= \del_{\cA} + i(\xag^{(0,1)})\;\;,\;\;\;\;\;\;     \label{qxag}
\bar{Q}(g)=\bar{\del}_{\cA} + i(\xag^{(1,0)})\;\;.
\ee
Within this context, the nilpotency of $Q$, $Q^{2} =0$, expresses the
holomorphicity of the vector field $\xag$ on the K\"ahler manifold $\cA$.
On $\xag$-invariant forms one also has $Q(g)\bar{Q}(g)=-\bar{Q}(g)Q(g)$.
Furthermore, on the fixed point set of $\xag$, $\d(g)$ reduces to the
ordinary exterior derivative.

It can be checked directly that $\xag$ is also a Killing vector field
(for the K\"ahler metric on $\cA$). Alternatively this follows from the
supersymmetry invariance of the action which implies that $\xag$ is
symplectic,
\be
\d(g)S(g,A,\p) = 0 \;\;\Ra\;\; i(\xag)\Omega(\p) =
d_{\cA}S_{G/G}(g,A)\;\;,
\ee
with Hamiltonian the $G/G$ action itself - see section 4. Hence,
since $\xag$ is holomorphic and $\cA$ K\"ahler, $\xag$ is Killing.
We are thus precisely in the setting of a K\"ahler manifold with a
holomorphic Killing vector field considered by Bismut \cite{bismut}
(albeit for a single vector field and not a whole family of them).

\subs{Splitting the Action of the $G/G$ Model}

Before explaining the relation between the $G/G$ model and Bismut's theory
of equivariant Bott-Chern currents, we will look at some more down-to-earth
consequences of the supersymmetry of the $G/G$ model.  These will eventually
lead us to the localization argument of the
next section. We will show that the $G/G$ action $S(g,A,\p)$ can be split into
a $Q\bar{Q}$-exact part and a cohomologically non-trivial piece.
This will allow us
to understand from a slightly different point of view the one-parameter family
of deformations of the $G/G$ model already considered by Witten in \cite{ew}.

We first rewrite the kinetic term $S_{G}(g,A)$ of the $G/G$ action as
\bea
S_{G}(g,A) &=& -\trac{1}{8\pi}\int_{\S}\g d_{A}g * \g d_{A}g\nonumber\\
           &=& -\trac{1}{4\pi}\int_{\S}\g D_{z}g\; \g D_{\zb} g\nonumber\\
           &=& -\trac{1}{4\pi}\int_{\S}J_{z}(g) (J_{\zb}(g))^{g}\;\;.
\eea
Since $\bar{Q}\p_{z}=J_{z}$ and $Q\p_{\zb}=J_{\zb}$, this term is actually
$Q\bar{Q}$-exact (modulo terms involving $\p$). The complete $G/G$ action
$S(g,A,\p)$ can, in fact, be written as
\be
S(g,A,\p) = -\trac{1}{4\pi}Q\bar{Q}\int_{\S}\p_{z}\p_{\zb}^{g} -\Gamma(g,A,\p)
\;\;,\label{split}
\ee
where
\be
\Gamma(g,A,\p) = i\Gamma(g,A) +
  \trac{1}{4\pi}\int_{\S}(\p_{z}\p_{\zb}+\p_{z}\p_{\zb}^{g})\;\;.\label{gamma}
\ee
The twisted symplectic form
\be
\Omega^{g}(\p) = \trac{1}{2\pi}\int_{\S}\p_{z}\p_{\zb}^{g}\label{omegag}
\ee
appearing in (\ref{split}) is equivariant,
\be
Q\bar{Q}\int_{\S}\p_{z}\p_{\zb}^{g}=-\bar{Q}Q\int_{\S}\p_{z}\p_{\zb}^{g}\;\;.
\label{qqom}
\ee
$\Omega^{g}(\p)$ is almost as natural a symplectic form to consider on
$\cA$ as $\Omega(\p)$. In particular, it is easily verified that
the vector field $\xag$ is also
Hamiltonian with respect to $\Omega^{g}(\p)$,
the corresponding Hamiltonian being
$-S_{G/G}(\g,A)$ instead of $S_{G/G}(g,A)$ for the untwisted symplectic form.

It follows from (\ref{qqom}) and
the supersymmetry of the action that $\Gamma(g,A,\p)$
is both $Q$- and $\bar{Q}$-closed (but not exact),
\be
Q \Gamma(g,A,\p) = \bar{Q} \Gamma(g,A,\p) = 0 \label{qg0}\;\;.
\ee
In particular, therefore, $\Gamma(g,A,\p)$ defines an equivariant cohomology
class on $\cA$. We will see later that it represents winding numbers or
Chern classes associated to reducible connections.

\subs{A One-Parmater Family of Deformations of the $G/G$ Model}

Since we have split the action of the $G/G$ model
into a $Q\bar{Q}$-exact piece and a rest, the theory should be independent
of the coefficient of the former which can then be used to localize the
functional integral. Let us therefore consider the one-parameter family
of theories given by
\be
S^{s}_{G/G}(g,A) =  s S_{G}(g,A) -i\Gamma(g,A) \;\;. \label{ggsact}
\ee
Actually, including the level $k$ (an integer) of the (gauged) WZW model,
we have a two-parameter family of theories, but $k$ will play no role in
the discussions of this section.

In \cite{ew}, Witten argued that classically this one-parameter family
of theories is constant as a variation with respect to $s$ is proportional
to the classical equations of motion $J(g)=0$ of the undeformed model.
Furthermore, the classical
equations of motion following from the variation of (\ref{ggsact}) are
equivalent to those of the undeformed model. In fact, varying $A_{z}$
and $A_{\zb}$ one finds
\bea
\g D_{z} g - \lambda D_{z}g\;\g&=&0\;\;,\\
D_{\zb}g\;\g - \lambda\g D_{\zb}g &=&0\;\;,
\eea
where
\be
\lambda = \frac{s-1}{s+1}\;\;.
\ee
For $0<s<\infty$ one has $-1<\lambda < 1$. Since $|\Ad(g)|\leq 1$ and the
equations of motion can be written as
\bea
(1-\lambda\Ad(\g))D_{z}(g) &=& 0\;\;,\\
(1-\lambda\Ad(g))D_{\zb}(g) &=& 0\;\;,
\eea
they are equivalent to the equations of motion $D_{z}g = D_{\zb}g = 0$ of the
$G/G$ model. Likewise the equation $F_{z\zb}=0$ is unaffected, as a variation
of $g$ in (\ref{ggsact}) leads to
\be
D_{\zb}(\g D_{z}g) + \lambda D_{z}(\g D_{\zb}g) + F_{\zb z} =0\;\;,
\ee
which, by $D_{z}g = D_{\zb}g = 0$, implies $F_{z\zb}=0$.

While this establishes the classical constancy of the one-parameter family
of theories $S_{G/G}^{s}$, Witten suggests that quantum mechanically the
invariance under $s\ra s+\d s$ is broken, because in the path integral the
change in $s$ can only be compensated by a field redefinition of $A$ which
involves $A$ itself, leading to a Jacobian which needs to be regularised.
Thus, quantum mechanically the $G/G$ model (for any value of $s$)
should be equivalent to the manifestly topological theory
at $s=0$, perturbed by quantum corrections of the kind
calculated in \cite{btver}. For the purposes
of localization we will be interested in the opposite limit $s\ra\infty$.

On the other hand, the supersymmetric extension of (\ref{ggsact}),
which will also
lead to an $s$ dependence of the term quadratic in the $\p$'s (see (\ref{spsi})
below), will automatically keep
track of these determinants. We will check below that, formally, the
ratio of determinants arising from the terms quadratic in $A$ and $\p$
is $s$-independent. Just as in \cite{btver}, their regularization
will give rise to quantum corrections to the $G/G$ action, in particular to
the shift $k\ra k + h$ of the level, confirming Witten's argument concerning
the relation between the theories for different values of $s$.

Consider now the supersymmetric extension of (\ref{ggsact}), given by
\be
S^{s}(g,A,\p) = -\frac{s}{4\pi}Q\bar{Q}
\int_{\S}\p_{z}\p_{\zb}^{g} -\Gamma(g,A,\p) \label{ssplit}
\;\;,
\ee
where $\Gamma(g,A,\p)$ was defined in (\ref{gamma}). The $\p$'s enter in
this action in the form
\be
\p_{z}[(1+s)+(1-s)\Ad(g)]\p_{\zb} = \p_{z}[(1+s)(1-\lambda
\Ad(g))]\p_{\zb}\;\;,
\label{spsi}
\ee
while the term quadratic in the gauge fields is
\be
\az[2+(1-s)\Ad(g)-(1+s) \Ad(\g)]\azb\;\;.
\ee
This can be factorized as
\be
               \az[((1+s)+(1-s)\Ad(g))(1-\Ad(\g))]\azb \;\;.
\ee
Thus formally the ratio of determinants is indeed $s$-independent and
given by the inverse square root of the determinant of the operator
$(1-\Ad(\g))$ acting on one-forms. This determinant, restricted to the
normal bundle of the fixed point locus, will arise upon localization
as the equivariant Euler class of the normal bundle, as in the
stationary phase formulae of Duistermaat-Heckman \cite{dh} and
Berline-Vergne \cite{bv}. It can also be checked that no $s$-dependence
is reintroduced into the action through the source terms coupling
to $A$ and $\p$.

\section{Localization of the $\bf G/G$ Model}

In this section we will show how the above considerations concerning
supersymmetry and deformations of the $G/G$ model can be used to
localize the $G/G$ functional integral and to therefore provide an
alternative derivation of the Verlinde formula in terms of equivariant
K\"ahler geometry. This localization could be carried out directly
on the basis of what we have established so far. Nevertheless we find it
interesting and instructive that precisely the structure of the $G/G$ model
described
in the previous section (the supersymmetry related to a holomorphic Killing
vector field, the action of the form $Q\bar{Q}$(symplectic form) plus a
cohomologically non-trivial piece) appears in the work of Bismut \cite{bismut}
on the relation between complex equivariant cohomology, Ray-Singer torsion,
Quillen metrics, and Bott-Chern currents. We therefore start with a brief
description of what we believe is the appropriate mathematical setting for
the $G/G$ model before working out the details of the localization. Ideally
this setting should allow one to establish directly that the $G/G$ action
represents the Riemann-Roch-Hirzebruch integrand for the Verlinde formula
in equivariant cohomology on $\cA$ but so far we have been unable to show
that.

\subs{The Mathematical Setting: Equivariant Bismut-Bott-Chern Currents}

We will have to introduce some notation.
Let $(M,\Omega)$ be a compact K\"ahler manifold and $X$ a holomorphic
Killing vector field on $M$ so that $L(X)\Omega = 0$. Denote by
$M_{X}$ the zero locus of $X$ (this is also a K\"ahler manifold),
by $N_{X}$ the normal bundle to $M_{X}$ in $M$ and by $J_{X}$ the
skew-adjoint endomorphism of $N$ given by the infinitesimal action of
$X$ in $N_{X}$. Let
$d_{X}=d + i(X)$ be the equivariant exterior derivative and $\del_{X}$
and $\bar{\del}_{X}$ the equivariant Dolbeault operators
\be
\del_{X} = \del + i(X^{(0,1)})\;\;,\;\;\;\;\;\;
\bar{\del}_{X} =  \bar{\del} + i(X^{(1,0)})\;\;, \label{bqxag}
\ee
satisfying the relations
\bea
&&(\del_{X})^{2}=(\bar{\del}_{X})^{2} = 0\;\;,\label{bqq0}\\
&&(\del_{X}+\bar{\del}_{X})^{2} =
  \del_{X}\bar{\del}_{X} + \bar{\del}_{X}\del_{X} = L(X)\;\;,\\
&& \del_{X}\bar{\del}_{X}\Omega = - \bar{\del}_{X}\del_{X}\Omega\;\;.
\label{bqqom}
\eea

In \cite{bismut}, Bismut studies integrals of the form
\be
\int_{M}\exp(-is\del_{X}\bar{\del}_{X}\Omega -i\Gamma) \label{bisint}
\ee
where $s$ is a real parameter and $\exp (-i\Gamma)$ is some smooth
(inhomogeneous) differential
form on $M$ which, in the cases of interest, is equivariantly closed with
respect to both $\del_{X}$ and $\bar{\del}_{X}$,
\be
\del_{X} \exp (-i\Gamma) = \bar{\del}_{X}\exp (-i\Gamma) = 0 \label{bqg0}\;\;.
\ee
These integrals are finite dimensional analogues of integrals which appear
in the loop space integral approach to index theorems and in the study
by Bismut, Gillet and Soul\'e \cite{bgs} of Quillen metrics on holomorphic
determinant bundles. In the infinite-dimensional case, $M$ would be the
loop space of a K\"ahler manifold and $X$ the
canonical vector field on the loop space
generating rigid rotations of the loops.
The equivariant cohomology of operators like $\del_{X}$ has been investigated
in \cite{vag}.

At this point  we can clarify the relationship of these considerations with
the formulae encountered in our discusion of the $G/G$ model:
$M$ is $\cA$, $X$ is
$\xag$, (\ref{bqxag}), (\ref{bqq0}) and (\ref{bqqom}) correspond to
(\ref{qxag}), (\ref{qq0}) and (\ref{qqom}) respectively, (\ref{bqg0})
is the counterpart of (\ref{qg0}), and the integral (\ref{bisint})
corresponds to the functional integral (over $\cA$) of the ($s$-deformed)
$G/G$ action (\ref{ssplit}).

As a consequence of (\ref{bqg0}), the integrand in (\ref{bisint}) has the
property that its $s$-derivative is
\be
\frac{\del}{\del s}\exp(-is\del_{X}\bar{\del}_{X}\Omega -i\Gamma)=
-\del_{X}\bar{\del}_{X}[i\Omega\exp(-is\del_{X}\bar{\del}_{X}\Omega -i\Gamma)]
\;\;, \label{bbc}
\ee
so that, in particular, the integral is $s$-independent. The first part
of the exponent can be written in the form
\bea
i\del_{X}\bar{\del}_{X}\Omega &=&
\frac{1}{2}d_{X}(\bar{\del}_{X}-\del_{X})i\Omega \nonumber \\
&=& \frac{1}{2}(d+i(X))X'\;\;,
\eea
where $X'$ is the metric dual of $X$. Hence one is in a position to apply
the standard
localization theorems of Duistermaat-Heckman \cite{dh} and Berline-Vergne
\cite{bv}. The essence of these theorems is that an equivariantly closed
form $\mu$, $(d + i(X))\mu = 0$ for $X$ a Killing vector field is equivariantly
exact away from the zeros of $X$.  To see this, define the (inhomogoenous)
differential form $\nu$ on the complement of a neighbourhood of $M_{X}$ in $M$
by
\be
\nu = \frac{\alpha}{1+d\alpha}\mu\;\;,
\ee
where $\alpha$ is the normalized metric dual of $X$,
\be
\alpha = X'/||X||^{2}\;\;.
\ee
As a consequence of the easily verified identities
\be
L(X)\alpha = i(X)d\alpha = 0
\ee
one finds that
\be
(d + i(X))\mu = 0 \Ra \mu = (d + i(X))\nu \;\;\mbox{on}\;\;M\setminus M_{X}
\;\;.
\ee
In particular, therefore, the top-form component of $\mu$ is exact on
$M\setminus M_{X}$ and the integral $\int_{M}\mu$ is determined by an
infinitesimal neighbourhood of $M_{X}$. Explicitly, the integral
(\ref{bisint}) is
\be
\int_{M}\exp(-is\del_{X}\bar{\del}_{X}\Omega -i\Gamma) =
\int_{M_{X}}\exp (-i\Gamma) E^{-1}(N_{X}) \;\;,\label{dh}
\ee
where $E(N_{X})$ is the equivariant Euler class of $N_{X}$, represented in
terms of $J_{X}$ and the curvature form $R(N_{X})$ of $N_{X}$ by
\be
E(N_{X}) = \det [\trac{i}{2\pi}(J_{X}+R(N_{X}))]\;\;.\label{E}
\ee
This can also be thought of as the square root of the determinant of the
operator acting on the underlying real bundle - a point of view
more natural in gauge theories.

In the $G/G$ model this formula can now be applied to (or derived from)
the functional integral over the gauge fields.
The main difference is, of course, that in the $G/G$ model
we are dealing with a family $\{\xag\}$ of holomorphic Killing vector fields,
indexed by $g\in\cal G$, as well as with a family of symplectic forms on $\cA$
(the twisted symplectic forms $\Omega^{g}(\p)$) with respect to
which the action takes the form (\ref{bisint}). Hence, for each
$g\in\cal G$ the gauge field functional integral will reduce to an integral
over the zero locus of $\xag$, i.e.~the connections satisfying $A^{g}=A$ and
this still needs to be integrated over $\cal G$.

Formulae like (\ref{bbc}) are very reminiscent of formulae characterizing
Bott-Chern forms (or currents). In fact,
Bott-Chern forms \cite{bc} are holomorphic analogues of the Chern-Simons
secondary characteristic classes of differential geometry. The latter
typically express the independence (in cohomology) of certain Chern-Weil
characteristic classes $\Phi^{CW}(F_{A})$ by transgression formulae like
\be
\Phi^{CW}(F_{A}) - \Phi^{CW}(F_{A'}) = d \Phi^{CS}(A,A')\;\;,
\ee
where $A$ and $A'$ are two connections on the same bundle. In the
holomorphic context one seeks analogous formulae with the exterior
derivative $d$ on the right hand side replaced by $\del\bar{\del}$ so
that one is dealing with a double transgression. For example, let
$E$ be a holomorphic vector bundle over a complex manifold $M$ and denote
by $\nabla^{h}$ the unique holomorphic Hermitian connection on $E$
associated to the Hermitian structure $h$ on $E$ and by $F_{h}$ its curvature.
Consider the (scaled) Chern character
\be
{\rm ch}(\nabla^{h}) = \Tr[\exp -(\nabla^{h})^{2}]\;\;.
\ee
Then the main results of Bott and Chern (see \cite{bc} and \cite{bgs})
are that under a variation of $h$ one has
\be
\d\Tr\exp [-(\nabla^{h})^{2}] =
        \del\bar{\del}\Tr [h^{-1}\d h\exp -(\nabla^{h})^{2}]
\ee
(this is to be regarded as the analogue of (\ref{bbc}))
and that this can be `integrated' to give an explicit expression for the
Bott-Chern class $\Phi^{BC}(h,h')$  satisfying
\be
\Tr[\exp -(\nabla^{h})^{2}] - \Tr[\exp -(\nabla^{h'})^{2}] = \del\bar{\del}
\Phi^{BC}(h,h')\;\;.
\ee

We hope that these analogies between the $G/G$ functional integral and
integrals of Bott-Chern currents will eventually lead to a better
cohomological understanding of the $G/G$ action.

\subs{Preliminary Remarks on Localization and the Fixed Point Locus}

It follows either from the above arguments (formally extended to
functional integrals) or from considering the $s\ra\infty$ limit
of the gauge field functional integral
\bea
Z_{G/G}(g,\p) &=& \int_{\cA}\! D[A]\,\exp [-S^{s}(g,A,\p)]\nonumber\\
&=& \int_{\cA}\! D[A]\,\exp [-\frac{s}{4\pi}Q\bar{Q}
\int_{\S}\p_{z}\p_{\zb}^{g} -\Gamma(g,A,\p)]\;\;, \label{aint}
\eea
that $Z_{G/G}(g,\psi)$ localizes onto the minima $\g d_{A}g =0$ of the
kinetic term, i.e.~onto the zero locus of $\xag$. A rough (but not quite
correct) path integral
argument for this would run roughly as follows. First one decomposes the gauge
field $A$ and and the group valued field $g$ into their `classical' and
`quantum' parts as
\be
A=A_{c}+A_{q}\;\;,\;\;\;\;\;\;g=g_{c}g_{q}\;\;,\;\;\;\;\;\;A_{c}^{g_{c}}=A_{c}
\;\;,\label{cq1}
\ee
and the quantum parts are taken to be orthogonal to the classical
configurations so that the quadratic form for the quantum fields is
non-degenerate. Then one scales the quantum fields by $1/\sqrt{s}$,
\be
A_{q} \ra A_{q}/\sqrt{s}\;\;,\;\;\;\;\;\;g_{q}=\exp\phi \ra \exp(\phi/\sqrt{s})
\;\;,
\ee
so that the quadratic term is $s$-independent. Then, in the limit $s\ra\infty$
only the determinant arising from the integral over the quantum fields and
the classical action $\Gamma(g_{c},A_{c})$ survive - which is just what
(\ref{dh}) expresses. Actually, in the case at hand we will have to be a
little bit more careful. The zero locus is still infinite-dimensional
and the quadratic form for $A_{c}$ is provided by part
of the quantum field $g_{q}$ (in fact, for fixed $g_{c}$, $\Gamma(g_{c},A_{c})$
turns out to be independent of $A_{c}$) which should hence not be scaled away.

The reason for the occurrence of this problem is the fact that the condition
for $A$ to be a critical point of the vector field $\xag$ is a condition on
both $A$ and $g$ while e.g.\ the localization theorem of the previous section
only applies (formally at least) to the $A$-part of the integral for fixed $g$.
Thus one should be careful to implement this $g$-dependent localization
correctly. This can be achieved by choosing a parametrization for $g$ in terms
of `classical' and `quantum' fields which is more explicit than the one used
above. In particular we will see that localization can be achieved by treating
the $g$-integral exactly, using the $s$-independence only to massage the
gauge field integral.

As this complication is already present in the large $k$ limit of the
$G/G$ model, i.e.\ in BF theory, and its origin as well as the way to handle
it are somewhat easier to understand in that example, we discuss it in
some detail in the appendix. Here we will instead present a streamlined version
of the argument adapted to the $G/G$ model.

Let us first take a closer look at the space
\be
\ag = \{A\in\cA: A^{g}=A\}
\ee
onto which the theory eventually localizes.
While usually the reducibility condition is regarded as an equation for
$g$ for a fixed $A$, here instead $g$ is fixed and one is looking for
gauge fields for which $g$ is contained in their isotropy group.
Nevertheless, this will turn out to be a condition on certain
components of $A$ as well as on $g$ since for most $g$ there will be no $A$
whatsoever satisfying $A^{g}=A$. For instance, by multiplying by powers
of $g$ and taking traces, one finds that
\be
A^{g}=A \Ra d\Tr g^{n} = 0 \;\;,\forall\;n\in\ZZ\;\;,
\ee
so that, essentially, $g$ is conjugate to a constant matrix.

In order to obtain some more information on $\ag$, we will use the method
of diagonalization introduced in \cite{btver} to calculate the $G/G$
functional integral. Thus assume that $g$ can be written as $g = ht h^{-1}$,
where $t\in\st$ takes values in the maximal torus $\bT$ of $\bG$.
Pointwise this can of course always be achieved, and the global issues
have been analyzed in detail in \cite{btdia}. In particular, one finds that
if $g$ is regular, i.e.~if at every point
$x\in\S$ the dimension of the centralizer of $g(x)$ in $\bG$ is equal to
the rank of $\bG$, then $t$ can be chosen to be smooth globally. Moreover,
the
torus component of the transformed gauge field $A^{h}$ is a connection on
a posibly non-trivial $\bT$ bundle over $\S$, indicating that $h$ will in
general not be smooth globally . The $\bT$ bundle in question
turns out to be \cite{btdia}
the pull-back of $\bG\ra\gt$ to $\S$ via the $\gt$-part of
a lift of $g$ to $\gt\times\bT$. Thus, for regular maps the reducibility
condition can be written as
\bea
A^{g}=A &\Leftrightarrow& A^{h} = A^{ht}\nonumber\\
        &\Leftrightarrow& (A^{h})^{\lt} = (A^{h})^{\lt} + t^{-1}dt      \\
                       && (A^{h})^{\lg/\lt} = t^{-1}(A^{h})^{\lg/\lt}t    \\
        &\Leftrightarrow& dt=0\;\;{\rm and}\;\; (A^{h})^{\lg/\lt}=0\;\;.
\eea
We therefore see that the localization essentially
abelianizes the theory and at this point the analysis can proceed more or less
as in \cite{btver}. In particular, for a regular $g$ with $h^{-1}g h = t$
constant, the space $\ag$ is isomorphic to the space of gauge fields on a
torus bundle over $\S$ and hence
\bea
dt \neq 0 &\Ra& \ag = \emptyset \nonumber\\
dt = 0    &\Ra& \ag \sim \Omega^{1}(\S,\lt)\;\;.
\eea
We want to draw attention to the fact that there is no
condition on the torus gauge field $(A^{h})^{\lt}$, so that that part of the
gauge field functional integral is not localized and needs to be calculated
directly.

At the other extreme, when $g$ is the identity matrix, there is no
localization at all, $\ag\sim\cA$, and the functional integral
(\ref{aint}) is hopelessly divergent as the action is then identically zero.
In general, some regularization prescription has to be adopted to deal with
highly non-regular elements of $\cal G$, for all of which the quadratic form
for the gauge fields in the $G/G$ action is in some sense degenerate. The
results of \cite{btver,btlec} suggest that any reasonable  prescription
should be tantamount to integrating only over regular
maps and discarding the non-regular maps. This is what we will henceforth do.

\subs{Evaluation of the Action on the Fixed Point Locus and Winding Numbers}

On $\ag$, the action $S_{G/G}^{s}(g,A)$  reduces to $-i\Gamma(g,A)$.
But, since
\be
\Gamma(g,A) = \Gamma(g)-\trac{1}{4\pi}\int_{\S} A dg\,\g + AA^{g}\;\;,
\ee
one finds that this simplifies to
\bea
\Gamma(g,A)|_{\ag} &=& \Gamma(g) - \trac{1}{4\pi}\int_{\S}A dg\,\g\nonumber\\
                   &=& \Gamma(g) - \trac{1}{4\pi}\int_{\S}A \g dg
                   \equiv W(g,A)
\eea
(where the second line follows from $A^{g}=A$ and $\int_{\S}(\g dg)^{2}=0$).
$W(g,A)$ is precisely the cocycle,
\be
W(gh,A) = W(g,A^{h^{-1}}) + W(h,A)\;\;, \label{wcoc}
\ee
implementing the lift of the $\cal G$ action to the prequantum line bundle
of Chern-Simons theory, i.e. to the line bundle over $\cA$ with curvature
form equal to ($k$ times) the basic symplectic form $\Omega(\p)$.

$W(g,A)$ has some more or less obvious properties which suggest that
it is a topological invariant associated with $g$. First of all, on $\ag$
it is of course invariant under smooth gauge transformations,
\be
A^{g}=A \;\;\Ra \;\; W(g^{h},A^{h}) = W(g,A)\;\;. \label{wgi}
\ee
What is more interesting, however, is that it is independent of $A\in\ag$,
\be
W(g,A) = W(g,A') \;\;\forall\;A,A'\in\ag\;\;.\label{ww}
\ee
The easiest way to see that is to use the representation $g = h t h^{-1}$
to write $A^{h}=a$ and $g^{h}=t$ where $a$ is a torus gauge fields and
$t$ is constant. Then
one finds
\bea
\int_{\S} A \g dg &=& \int_{\S}(hah^{-1} -
dh\,h^{-1})(ht^{-1}h^{-1}dh t h^{-1} -dh\,h^{-1})\nonumber\\
&=& \int_{\S} t^{-1} h^{-1} dh t h^{-1} dh\;\;.\label{agdg}
\eea
As this is independent of $A$, the claim follows. Hence, as mentioned above,
a quadratic form for the $A_{c}$-integration will have to be provided by
those parts of $g_{q}$ which couple to $A_{c}$. We will discuss this in more
detail below.

Another property of
$W(g,A)$, which follows directly from the cocycle identity (\ref{wcoc}),
using $A^{g}=A$, is
\be
W(g^{n},A) = n W(g,A) \;\;\forall\;n\in\ZZ\;\;.
\ee
These observations taken together strongly suggest that $W(g,A)$ is
related to the winding numbers for regular maps $g\in\sgr$ introduced
in \cite{btdia}. These winding numbers and winding number sectors
exist in $\sgr$ because, although the non-regular elements of $\bG$ are
of codimension three in $\bG$, maps from a two-manifold into $\bG$ which
pass through a non-regular element somewhere are actually of codimension
one. Hence, in contrast to $\sg$,
$\sgr$ is not connected but turns out to consist of a $\ZZ^{r}$'s
worth of connected components, where $r$ is the rank of $\bG$,
\be
\pi_{0}(\sgr) = \ZZ^{r}\;\;.
\ee
It is no coincidence that this is the same as $\pi_{2}(\gt)$. Explicitly,
these winding numbers of $g$ can be written as
\be
n^{l}(g) = \trac{1}{4\pi}\int_{\S} \alpha^{l} [h^{-1}dh,h^{-1}dh]
\;\;\;\;\;\;l = 1,\ldots,r\;\;,\label{wn}
\ee
where the $\alpha^{l}$ are simple roots of $\bG$.
To see how they are
related to $W(g,A)$, let us write $t$ as
$t = \exp \phi$ where $\phi =\alpha^{l}\phi_{l}$ is constant.
Then the relationship between $W(g,A)$ and $n^{k}(g)$ is
\be
W(g,A) = \phi_{l} n^{l}(g)\;\;.\label{w=n}
\ee
Thus on the fixed point set $\ag$, the action of the $G/G$ model simply
reduces to a linear combination of the winding numbers (\ref{wn}),
\be
kS_{G/G}(g,A)|_{\ag} = ik\phi_{l}n^{l}(g)\;\;. \label{wnt}
\ee
These winding numbers are
also the Chern classes of the connection $(A^{h})^{\lt}$
\cite[Corollary 4]{btdia}. We will obtain both these results in the next
section when discussing how the action of the $G/G$ model reduces to that
on the fixed point locus analyzed here.

\subs{Path Integral Derivation of the Localization}

Having analyzed the `classical' action of the $G/G$ model, we will still have
to establish how and in which sense localization reduces the path integral
to an integral over the classical fields.
Above we sketched a rough (albeit wrong)
path integral argument for localization in the $G/G$ model. We will now
present a more careful argument which has the virtue of being correct. It is
the exact counterpart of the method (more precisely, method (2))
used in the appendix for solving Yang-Mills
theory, and we refer to the appendix for a more detailed discussion in that
case.

We begin by writing the action of the (deformed)
$G/G$ model more explicitly in terms
of the `classical' and `quantum' fields.
It follows from the above that a general regular map $g$ can be written in the
form
\be
g = ht_{c}\bar{t}h^{-1}= (ht_{c}h^{-1})(h\bar{t}h^{-1})\;\;,\label{cqdec}
\ee
where the classical field $t_{c}$ is constant
and $\bar{t}$ contains no constant mode.
Note that (\ref{cqdec}) is invariant under $h\ra h \tau$ for $\tau\in\st$,
while $h\ra \gamma h$ for $\gamma\in\sg$ generates the adjoint (gauge)
transformation on $g$. The second equality in (\ref{cqdec}) represents
the improved and refined version of the decomposition $g=g_{c}g_{q}$
used in (\ref{cq1}). If we plug this form of $g$ into the deformed bosonic
$G/G$ action (\ref{ggsact}), it is clear that by gauge invariance the first
term can alternatively be written as
\be
S_{G}(g,A) = S_{G}(t_{c}\bar{t},\ah)\;\;.\label{cqact}
\ee
Clearly, $\ah$ is invariant under gauge transformations. It is, however,
not invariant under the `parametrization symmetry' $h\ra h\tau$. On the
other hand, this is certainly an invariance of the action as the fields
appaearing on the left hand side of (\ref{cqact}) are inert under this
transformation.

Decomposing $\ah$ into its $\lt$- and $\lg/\lt$-components,
$\ah=\aht + \ahk$, we can see what $h\ra h\tau$ implies for $\ah$
explicitly:
\bea
&&\aht\ra\aht + \tau^{-1}d\tau\;\;,\nonumber\\
&&\ahk\ra \tau^{-1}\ahk\tau\;\;.
\eea
Thus $\aht$ transforms as and hence is a connection on some $\bT$ bundle,
while the $\lg/\lt$-component is a section of a bundle associated to it via
the adjoint action of $\bT$ on $\lg/\lt$.

In terms of this decomposition of $\ah$, the action (\ref{cqact}) becomes
\be
S_{G}(g,A)= -\trac{1}{8\pi}\int_{\S} d\bar{t}*d\bar{t} +
(1-\Ad(t_{c}\bar{t}))\ahk * (1-\Ad(t_{c}\bar{t}))\ahk \;\;.  \label{sgcq}
\ee
The gauged WZ term requires a little bit more care. Technically the reason for
this is that, in contrast to $S_{G}(g,A)$, $\Gamma(g,A)$ is not invariant
under arbitrary, possibly discontinuous, gauge transformations, the integrand
transforming homogenously under gauge transformations only up to a
total derivative on $\S$. Hence one cannot invoke gauge invariance to (falsely)
conclude that $\Gamma(g,A) = \Gamma(t_{c}\bar{t},\ah)$ because $h$ may not be
continuous. We will instead calculate $\Gamma(ht_{c}\bar{t}h^{-1},A)$ directly.

Let us start with the WZ term $\Gamma(g)$. We have to find an
extension of $g=ht_{c}\bar{t}h^{-1}$ to some bounding three-manifold $N$.
First of all we choose $N$ to be $N = \S\times [0,1]$ with
$\del N = \S\times\{1\}-(\S\times\{0\})$. We will now extend $g$ to $N$
in such a way that $g|_{\S\times\{0\}}=1$ so that there are no contributions
to the action from that part of the boundary. Writing $t=t_{c}\bar{t}$
as  $t_{c}\bar{t} = \exp \f \equiv \exp (\f_{c} + \fb)$, we choose this
extension to be simply
\be
g(x,s) = h(x) \exp (s\phi) h^{-1}(x)\;\;,
\ee
which has the desired properties
\be
g(x,0) = 1\;\;,\;\;\;\;\;\;g(x,1) = g(x)\;\;,
\ee
as well as preservation of the right $\bT$-invariance of $h$.
It is then a matter of straightforward calculation to determine $\Gamma(g)$:
\bea
\Gamma(g) &=&
\trac{1}{12\pi}\int_{\S}\int_{0}^{1}\!ds\,(g(x,s)^{-1}dg(x,s))^{3}\nonumber\\
&=&\trac{1}{4\pi}\int_{\S}\phi [h^{-1}dh,h^{-1}dh] \nonumber\\
&+&\trac{1}{4\pi}\int_{\S}\int_{0}^{1}\!ds\,\frac{d}{ds}
  [\exp (-s\phi) h^{-1} dh \exp (s\phi) h^{-1}dh ]\nonumber\\
&=& (\f_{c})_{l} n^{l}(g) +
\trac{1}{4\pi}\int_{\S}\fb[h^{-1}dh,h^{-1}dh] +
\trac{1}{4\pi}\int_{\S}t^{-1} h^{-1} dh t h^{-1} dh
\;\;.
\eea
We see here the emergence of the winding number term (\ref{wn},\ref{w=n})
anticipated in the previous section. Determining the remaining terms of
$\Gamma(g,A)$ is straightforward and, putting everything together, the
gauged WZ term can be written as
\bea
\Gamma(g,A) &=& \trac{1}{4\pi}\int_{\S}\f [h^{-1}dh,h^{-1}dh] +
              \trac{1}{2\pi}\int_{\S}h^{-1}dh\,d\fb \nonumber\\
            &-& \trac{1}{4\pi}\int_{\S}\ahk\Ad(t_{c}\bar{t})\ahk
               -\trac{1}{2\pi}\int_{\S}\aht d\fb \;\;. \label{gammacq}
\eea
This expression is not yet particularly transparent. In particular,
as $\aht$ may be a connection on a non-trivial torus bundle, it is
not clear that (\ref{gammacq}) is even well defined. However, because
of the interplay between winding numbers and Chern classes this is
indeed the case. In particular, although integrating by parts the last term
is illegal, the first, second and fourth terms combine to give
\bea
\Gamma(g,A) &=& -\trac{1}{2\pi}\int_{\S}\f F_{\aht} -\trac{1}{4\pi}\int_{\S}
\ahk\Ad(t_{c}\bar{t})\ahk\label{gammacq1}\\
&-& \trac{1}{2\pi}\int_{\S}d (\fb h^{-1} A h)\nonumber\;\;.
\eea
This makes it manifest that (\ref{gammacq}) is globally well defined.
The last term could be non-zero only because of possible winding modes of
$\fb$ but we will see immediately that this term is actually not there at all.

One can now decompose $\aht$ into the sum of a background connection $A_{0}$
and a one-form $a^{\lt}$. Integrating over the latter imposes the condition
that $d\fb=0$ and hence $\fb=0$ as $\fb$ has no constant modes. Hence
$\bar{t}$ disappears from $S_{G}(g,A)$  while
(\ref{gammacq}) and (\ref{gammacq1}) reduce to
\be
\Gamma(g,A) \ra  \trac{1}{4\pi}\int_{\S}\f_{c}[h^{-1}dh,h^{-1}dh]
                    -\trac{1}{4\pi}\int_{\S}\ahk\Ad(t_{c})\ahk
\ee
and
\be
\Gamma(g,A) \ra -\trac{1}{2\pi} \Tr \f_{c}\int_{\S}F_{A_{0}}
                    -\trac{1}{4\pi}\int_{\S}\ahk\Ad(t_{c})\ahk  \label{gammaf}
\ee
respectively. This establishes among other things the relation between
the winding numbers $n^{l}(g)$ of (\ref{wn}) and the Chern classes of the
corresponding torus connection $\aht$,
\be
n^{l}(g) = -\trac{1}{2\pi}\int_{\S}\alpha^{l}F_{\aht} \equiv n^{l}(\aht)\;\;.
\ee

There are now several ways to calculate the path integral over the remaining
fields (and hence the partition function of the $G/G$ model). One possibility
is to choose the gauge $h=1$.
This is essentially what we did in \cite{btver} and is the Abelianization
approach to the evaluation of the path integral which we will not repeat here.
We just mention that for $s=1$ the terms from
(\ref{sgcq}) and (\ref{gammacq1}) involving $\ahk=A^{\lk}$ combine to give the
chiral quadratic form $A_{z}^{\lk}(1-\Ad(t^{-1}))A_{\zb}^{\lk}$ whose
determinant formally cancels against the Faddev-Popov determinant up to
zero modes.

Alternatively one can now solve the theory via localization. To that end
we scale $\ahk$ and its superpartner by
\be
\ahk \ra s^{-1/2} \ahk \;\;,\;\;\;\;\;\;(\p^{h})^{\lk}\ra s^{-1/2}
(\p^{h})^{\lk}\;\;,
\ee
(as the $\p$'s are Grassmann odd, this does not introduce any $s$-dependence
in the measure)
and take the limit $s\ra\infty$. In this limit, (\ref{sgcq}) reduces
to
\be
-\trac{1}{4\pi}\int_{\S}(\ah_{z})^{\lk}(1-\Ad(t_{c})^{2})(\ah_{\zb})^{\lk}
\;\;,
\ee
only the first (topological) term of (\ref{gammaf}) survives, and
the fermionic part (\ref{spsi}) becomes
\be
-\trac{1}{4\pi}\int_{\S} (\p_{z}^{h})^{\lk}(1-\Ad(t_{c})(\p_{\zb}^{h})^{\lk}
-\trac{1}{4\pi}\int_{\S}(\p^{h})^{\lt}(\p^{h})^{\lt} \;\;.
\ee
Note that the action is still completely gauge invariant under
\be
A\ra A^{\tilde{g}}\;\;,\;\;\;\;\;\;h\ra \tilde{g}^{-1} h
\ee
(the latter corresponding to $g\ra \tilde{g}^{-1} g \tilde{g}$), as the gauge
fields only appear in the manifestly gauge invariant St\"uckelberg combination
$\ah$).

Changing variables from $A$ and $\p$ to $\ah$ and $\p^{h}$ the $h$-integral
(gauge volume) factors out. Performing
the integral over $A^{\lk}$ and $\p^{\lk}$, one obtains the functional
determinant
\be
\Det^{-1/2}[1-\Ad(t_{c})]|_{\Omega^{1}(\S,\lg/\lt)}\;\;.\label{detone}
\ee
Here we have indicated explicitly that this is a functional determinant
on the space of $\lk$-valued one-forms on $\S$.
On the other hand, the Jacobian from the change of variables $g\ra(h,t)$ is
\be
\Det [1-\Ad(t)]|_{\Omega^{0}(\S,\lg/\lt)}\;\;. \label{jacobian}
\ee
Obviously these determinants almost cancel. It has been shown in
\cite{btver,btlec} that
this ratio is (up to a phase) a finite-dimensional determinant,
arising from the unmatched harmonic modes between zero- and one-forms.
Using a regularization that preserves the $\bT$ gauge symmetry, one explicitly
finds that the regularized product of (\ref{detone})
and (\ref{jacobian}) is
\be
\exp (ih\phi_{l}n^{l}(g))
\det{}^{\chi(\S)/2} (1-\Ad(t_{c}))|_{\lg/\lt}\;\;,\label{ratio}
\ee
where $h$ is the dual Coxeter number of $\bG$ and $\chi(\S)$ the Euler number
of $\S$. The remaining action
is then simply the linear combination of Chern classes appearing in
(\ref{wnt}) or (\ref{gammaf}), i.e.\
the $G/G$ action evaluated on the classical
configurations (the zero locus of the vector field $\xag$) $A^{g}=A$,
the net effect of the phase in (\ref{ratio}) being to shift the
coefficient of this term from the level $k$ to $k+h$.

Modulo the modifications brought about the fact that the $G/G$ partition
function includes an integral over $g$, this result agrees exactly
with the predictions of Bismut's localization formula (\ref{dh})
for the integral (\ref{bisint}). For example,
it is easy to see that the determinant (\ref{detone}) is precisely
the equivariant Euler class (\ref{E}) of the normal bundle to $\ag$
appearing in (\ref{dh}). The only thing to note is that
the normal bundle $\ng$ to $\ag$ in $\cA$ is trivial, as
$\ag$ is contractible (it need of course not be equivariantly trivial) and that
$\ng$ has also got vanishing curvature in terms
of the connection it inherits from the (flat) Levi-Civita connection
on $\ag$. Thus the equivariant curvature $J_{X}+R(N_{X})$ is given entirely
by the scalar part $J_{X}\sim J(g)$ which acts as
\be
J(g)Y = (Y_{z}^{g}-Y_{z})dz + (Y_{\zb}-Y_{\zb}^{\g})d\zb \;\;,
\ee
leading to the above determinants.

\subs{Putting Everything Together: The Verlinde Formula}

One can now follow exactly the same steps as in \cite[section 7]{btver}
to complete the evaluation of the partition function.
As the path integral derivation of the Verlinde formula has been
explained in detail in \cite{btver,btlec}, we will be rather brief
in this section. We will collect the results obtained above and then
only summarize the main steps of the evaluation. The interested reader is
referred to  \cite{btver,btlec}.

As a consequence of what we have learnt so far, we already know that the
$G/G$ partition function
\be
Z_{G/G}(\S,k)=\int_{\cal G}\! D[g]\int_{\cA}\! D[A] \exp(-kS_{G/G}(g,A))
\ee
reduces
to an expression involving only an infinite sum (arising from the sum over
all isomorphism classes of torus bundles on $\S$) and a finite dimensional
integral over $\bT$,
\be
Z_{G/G}(\S,k)=\sum_{(n^{l})\in\ZZ^{r}}\int_{\bT}\!\prod_{l=1}^{r}d\phi^{l}\,
                 \exp(i(k+h)\phi_{l}n^{l})\det{}^{\chi(\S)/2}
                 (1-\Ad(\exp\alpha^{l}\phi_{l}))|_{\lg/\lt}\;\;.
\ee
Now the infinite sum is a periodic delta function giving rise to a quantization
condition on the torus fields $\phi_{l}$. The allowed values of $\phi_{l}$
are
\be
\sum_{(n^{l})}\;\;\Ra\;\; \phi_{l} = \frac{2\pi m_{l}}{k+h}\;\;,\;\;\;\;
m_{l}\in\ZZ\;\;.
\ee
This turns the integral over $\bT$ itself into a sum. As the $\phi_{l}$
are compact scalar fields, only a finite number of the discrete values for
$\phi$ are allowed and hence this sum is finite. By restricing the sum to be
over regular elements of $\bT$ only and by eliminating the residual
Weyl group invariance, this sum can be shown to be a sum over the
integrable representations of the group $\bG$ at level $k$. For example, for
$\bG=SU(n)$ one finds
\be
\phi_{l} = \frac{2\pi m_{l}}{k+n}\;\;,\;\;\;\;m_{l}>0\;\;,\;\;
\sum m_{l} < k+h \label{phiq}
\ee
(the values $0$ and $k+h$ have been excluded because they
correspond to non-regular values of $t$). The range of the $m_{l}$ is
precisely the range labelling the integrable representations of $SU(n)$.
To be even more concrete, let us consider the case $\bG=SU(2)$. Using
$\det(1-\Ad(t))\propto\sin^{2}\phi/2$, one finds that
\be
Z_{SU(2)/SU(2)}(\S,k) \propto
\sum_{l=1}^{k+1}\left(\sin\frac{\pi l}{k+2}\right)^{\chi(\S)}\;\;.
\ee
Up to a normalization factor $((k+2)/2)^{\chi(\S)}$ (which can also be
determined - see \cite{btver}) this is indeed the $SU(2)$ Verlinde formula.
Analogously one obtains the Verlinde formula for other compact groups.
We refer to \cite{btver,btlec} for further details
concerning e.g.~the range of $\phi_{l}$ and the role of the action of
the Weyl group and to \cite{btdia,szenes} for what happens in the case of
non-simply connected groups and/or non-trivial $\bG$-bundles.

\section{The $\bf G/G$ Action as a Generalized Moment Map}

In this section we want to uncover the symplectic geometry underlying the
supersymmetry of the $G/G$ model. As already hinted at above, the structure
that we will find is not that of ordinary Hamiltonian group actions on
symplectic manifolds together with their infinitesimal moment maps but rather a
globalization thereof in which the role played by the Lie algebra in the
usual setting is played by the group (or rather, as we will see, by its
group algebra) instead. We will find that the $G/G$ action can
be interpreted as such a generalized moment map for the group action on $\cA$
generated by the vector fields $\xag$. Furthermore the (generalized)
equivariance of this moment map turns out to be equivalent to the
Polyakov-Wiegmann identity and
hence determines the $A$-independent part of the action to be the WZW action
$S_{G}(g)$.

\subs{A Brief Review of Hamiltonian Group Actions}

Let $(M,\Omega)$ be a symplectic manifold and $\bH$ be a group acting by
diffeomorphisms on $M$. Denote by $X_{M}$ the vector field on $M$ corresponding
to $X\in\lh = {\rm Lie}\bH$ so that one has
\be
[X_{M},Y_{M}]=[X,Y]_{M}\;\;. \label{xy}
\ee
The action on $M$ is said to be symplectic if each vector field $X_{M}$
leaves the symplectic form invariant,
\be
L(X_{M})\Omega \equiv (d+i(X_{M}))^{2}\Omega = 0\;\;\;\;\;\;\forall \;X\in
\lh\;\;.
\ee
As $\Omega$ is closed this is eqivalent to $di(X_{M})\Omega =0$. If
$i(X_{M})\Omega$ is not only closed but actually exact,
\be
i(X_{M})\Omega = dF(X) \label{sg1}
\ee
for some function $F(X)$ on $M$, then the action is said to be
Hamiltonian with $X_{M}\equiv V_{F(X)}$ the corresponding Hamiltonian
vector field. Note that this defines $F(X)$ only up to the addition of an
$X$-dependent constant $c(X)$.
It follows that the inhomogenous form $F(X)-\Omega$ is equivariantly closed,
\be
(d+i(V_{F(X)}))(F(X)-\Omega) = 0\;\;.
\ee
Introducing a basis $\{X_{a}\}$ of $\lh$ such that $[X_{a},X_{b}]
= f_{ab}^{c} X_{c}$ and $X=\phi^{a}X_{a}$,
and denoting the corresponding Hamiltonian and Hamiltonian vector field by
$F_{a}$ and $V_{a}$ respectively, this can also be written as
\be
(d+\phi^{a}i(V_{a})) (\phi^{a}F_{a}-\Omega) = 0\;\;.\label{sympact}
\ee
The operator on the left hand side is nilpotent on $\bH$-invariant forms
and its cohomology can be used to define the $\bH$-equivariant cohomology
$H^{*}_{\bH}(M)$ of $M$. For more on the relation betwen this (Cartan)
and other models of equivariant cohomology see \cite{jaap}.

The collection of functions $\{F(X)\}$ can equivalently be thought of as
either a map from $\lh$ to $C^{\infty}(M)$ or as a map $J$
from $M$ to the dual $\lh^{*}$ of the Lie algebra of $\bH$. These
two pictures are related by
\bea
&&F:\lh\ra C^{\infty}(M)\label{mom1}\\
&&J:M\ra \lh^{*}\label{mom2}\\
&&J(m)(X) = F(X)(m)\;\;.\label{mom3}
\eea
$J$ is called the moment map of the Hamiltonian group action.

If one defines the Poisson bracket of two functions $F(X)$ and $F(Y)$
in the usual way by
\be
\{F(X),F(Y)\} = L(V_{F(X)})F(Y) = i(V_{F(X)})i(V_{F(Y)})\Omega\;\;,
\ee
then it follows straightforwardly from the definitions that
\be
V_{\{F(X),F(Y)\}} = [V_{F(X)},V_{F(Y)}]=V_{F([X,Y])}\;\;.\label{equiv1}
\ee
Because of the non-degeneracy of the symplectic form this implies that
\be
d\{F(X),F(Y)\} = d F([X,Y])\;\;,
\ee
Hence the Poisson bracket $\{F(X),F(Y)\}$ differs from
the Hamiltonian $F([X,Y])$ only by a constant $c(X,Y)$. As a consequence of
the Jacobi identity, $c(.,.)$ defines a two-cocycle on $\lh$.
If this two-cocylce is trivial, the constants $c(X)$ can be adjusted
in such a way that
\be
\{F(X),F(Y)\}=F([X,Y]) \;\;\;\;\;\;\forall\; X,Y \in \lh\;\;,
\label{equiv2}
\ee
i.e.~such that
\be
\{F_{a},F_{b}\} = f_{ab}^{c}F_{c}\;\;.
\ee
Then the assignment
$X\ra F(X)$ defines a Lie algebra morphism from $\lh$ to
the (Poisson) Lie algebra $C^{\infty}(M,\Omega)$. In that case the
moment map $J$ intertwines the $\bH$-action on $M$ and the coadjoint
action on $\lh^{*}$ and is said to be equivariant. If either the second
Lie algebra cohomology group of $\lh$ is trivial,
$H^{2}(\lh)=0$, or $M$ is compact, equivariance can always
be achieved (in the latter case one can fix the constants $c(X)$ by
demanding that $\int_{M}\!d\mu\, F(X) =0$ where $d\mu$ is the Liouville
measure on $M$). Furthermore, if $H^{1}(\lh)=0$ (i.e.~if $[\lh,\lh]
=\lh$), equivariance fixes the moment map uniquely (otherwise one can,
without violating (\ref{equiv2}),
add any functional $c(.)$ to $F$ which vanishes on commutators, i.e.
$c\in (\lh/[\lh,\lh])^{*}$).

This sort of structure occurs naturally in e.g.~2d Yang-Mills theory
or BF theory.
The moment map is the generator of gauge transformations on the
symplectic space $\cA$ of 2d connections and the action is of the form
$\phi^{a}F_{a}-\Omega$ (plus a term quadratic in $\phi$ for Yang-Mills theory),
where now $\phi^{a}$ is a Lie algebra valued scalar field and $F_{a}$ is the
curvature two-form. In this context (\ref{sympact}) expresses
the (equivariant) supersymmetry of the theory - see \cite{ew2d}
or \cite{btlec} for more information.

\subs{Interpretation of the $G/G$ Action as a Generalized Moment Map}

The structure one finds in the $G/G$ model is to a large extent analogous
to the one discussed above, the main difference being that the infinitesimal
group action on $\cA$ is not paramterized by elements of the Lie algebra of
the gauge group $\cal G$ but rather (see (\ref{xag},\ref{dxag})) by the
elements of
$\cal G$ itself. This then presents a departure form the standard theory
of Hamiltonian group actions and some of the concepts (like the equivariance
condition) will have to be modified accordingly. We will show first how this
structure arises in the $G/G$ model and then extract from it the general
features in analogy with what we did above in the case of ordinary
Hamiltonian group actions.

As a first step we rewrite the supersymmetry $\d S(g,A,\p)=0$ of the action
(\ref{susact}) as
\be
i(\xag)\Omega(\p) = d_{\cA}S_{G/G}(g,A)\;\;, \label{iods}
\ee
where $2\pi\Omega(\p) =\int_{\S}\p_{z}\p_{\zb}$ denotes the symplectic
form on $\cA$. Comparing this with (\ref{sg1}) we are tempted to interpret
the action of the $G/G$ model as the `moment map' for the action generated
by $\xag$ on $\cA$, the crucial difference being that this moment map now
depends non-linearly on $g\in\bH=\cal G$ rather than linearly on
$X\in\lh$. There exists an exact analogue of the first description
(\ref{mom1}) of the moment map by regarding $F(g)\equiv S_{G/G}(g,.)$ as
a function(al) on $\cA$,
\bea
&&F:\bH={\cal G}\ra C^{\infty}(\cA)\label{gmom1}\\
&&F(g)(A) = S_{G/G}(g,A)\label{gmom3}\;\;.
\eea
The counterpart of the second description (the moment map $J$ as a map from
the symplectic manifold to the space $\lh^{*}$ of linear functions on  $\lh$)
can be obtained by replacing $\lh^{*}$ by
a space ${\cal F}({\cal G})$ of function(al)s on $\cal G$,
\bea
&&J:\cA\ra {\cal F}({\cal G}) \label{gmom2}\\
&&J(A)(g)=S_{G/G}(g,A)\;\;.
\eea
Infinitesimally, of course, these correspond to linear functions on
${\rm Lie}\cal G$, $F$ inducing a linear map
$T_{1}F$ (the derivative at the identity element) from ${\rm Lie} \cal G$ to
$C^{\infty}(\cA)$.

\subs{Global Equivariance of the Moment Map and the Polyakov-Wiegmann
Identity}

One other thing worth noting about (\ref{iods}) is that it only determines
the $A$-dependent part $S_{/G}(g,A)$ of $S_{G/G}(g,A)$ and that any
functional of the form $F(g)=S_{/G}(g,.) + C(g)$ will also satisfy
(\ref{iods}).
This is the analogue of the ambuiguity $F(X)\ra F(X)+c(X)$ we discussed above.
There this ambiguity could be fixed by demanding equivariance of the moment
map. It is thus natural to ask whether a similar criterion can be used here
to determine $C(g)$ to be the WZW action $S_{G}(g)$. This turns out to be
the case. To get an idea of what the analogue of the equivariance condition
(\ref{equiv2}) should be, we shall first determine the counterpart of
(\ref{equiv1}) and then try to lift that to an equation at the level of
moment maps.

By straightforward calculation one finds that the Lie bracket of two
vector fields $\xag$ and $\xa(h)$ is
\be
[\xa(g),\xa(h)]=\xa(gh)-\xa(hg)\;\;.\label{gequiv1}
\ee
Hence the equivariance condition one expects the moment map to satisfy
is
\be
\{F(g),F(h)\}=F(gh)-F(hg)\label{gequiv2}\;\;,
\ee
which is the (not completely obvious, but natural) counterpart of
$\{F(X),F(Y)\}=F([X,Y])$ (\ref{equiv2}). The interpretation of
this equivariance condition and its relation with (\ref{equiv2}) will be
discussed below. We will now show that with
the choice $F(g)(A)=S_{G/G}(g,A)$ (i.e.~with $C(g)=S_{G}(g)$) this
equation is satisfied.

It follows from the generalized Polyakov-Wiegmann identity (\ref{pw2})
that the right hand side of (\ref{gequiv2}) is
\be
S_{G/G}(gh,A)-S_{G/G}(hg,A) =
     \trac{1}{2\pi}\int_{\S}J_{z}(h)J_{\zb}(g)-J_{z}(g)J_{\zb}(h)\;\;.
\ee
The Poisson bracket on the left hand side can be calculated by contracting
$\Omega(\p)$ with $\xa(g)$ and $\xa(h)$ and one finds
\bea
\{S_{G/G}(g,A),S_{G/G}(h,A)\} &=& i(\xag)i(\xa(h))\Omega(\p)\nonumber\\
&=&\trac{1}{2\pi}i(\xag)\int_{\S}J_{z}(h)\p_{\zb}-J_{\zb}(h)\p_{z}\nonumber\\
&=&\trac{1}{2\pi}\int_{\S}J_{z}(h)J_{\zb}(g)-J_{z}(g)J_{\zb}(h)\;\;,
\eea
so that one indeed has the rather remarkable equation
\be
\{S_{G/G}(g,A),S_{G/G}(h,A)\} = S_{G/G}(gh,A)-S_{G/G}(hg,A)  \label{gh}
\ee
satisfied by the $G/G$ action with respect to the Poisson bracket on the
space $\cA$ of gauge fields.

However, demanding that (\ref{gequiv2}) holds still does not fix $F(g)$
uniquely to be the $G/G$ action. It is clear from the above that any
functional of the form $S_{G/G}(g,.) + C(g)$ whith $C(.)$ a class function
will also satisfy (\ref{gequiv2}), as then $C(gh)=C(hg)\;\forall\;g,h$.
In particular, this allows us to add to the $G/G$ action any of the observables
of the $G/G$ model like traces of $g$ as well as possible quantum corrections
which are also of this form \cite{btver,ew} without loosing the underlying
equivariant geometry.

\subs{Deformations of the Generator of Gauge Transformations and the
$k\ra\infty$ Limit}

We now want to show that the level $k$ $G/G$ action and its equivariance
relation (\ref{gh}) can be regarded as a deformation of the ordinary generator
of gauge transformations on $\cA$, the BF action
\be
S_{BF}(\phi,A) = \trac{1}{2\pi}\int_{\S}\phi F_{A}
\ee
(with $\phi\in{\rm Lie}\cal G$), and its standard equivariance condition
\be
\{S_{BF}(\phi,A),S_{BF}(\phi',A)\} = S_{BF}([\phi,\phi'],A)\;\;,\label{ff}
\ee
to which it reduces in the $k\ra\infty$ limit. That the level $k$
plays the role of a deformation parameter in the $G/G$ model can
also be seen form several other points of view. For instance, while
the partition function of BF (and Yang-Mills) theory is given by
a sum over all unitary irreps of $\bG$, in the $G/G$ model at level $k$
only the level $k$ integrable representations appear so that the finiteness
of $k$ effectively provides a cutoff on the representations contributing
to the partition function. These integrable representations are also known
to be related to the representations of quantum groups $\bG_{q}$ for
$q$ a root of unity, $q = \exp (i\pi/k+h)$, with $q\ra 1$ for $k\ra\infty$.
This suggests that the correct cohomological interpretation of the
supersymmetry and the localization could be in terms of (a yet to be
developed) $\bG_{q}$ (or rather ${\cal G}_{q}={\rm Map}(\S,\bG_{q})$)
equivariant cohomology on the space of gauge fields.

Moreover, it follows from
the Riemann-Roch formula (or from standard arguments concerning the
semi-classical limit of a quantum theory) that the large $k$ limit of
the Verlinde formula, or of the partition function of the $G/G$ model,
calculates the volume of the moduli space of flat connections, in
agreement with the fact that this is what the BF theory calculates.

To keep track of the $k$-dependence, we rewrite (\ref{gh}) in terms
of the $G/G$ action at level $k$ and the Poisson bracket
\be
\{.,.\}_{k} = k^{-1}\{.,.\}_{k=1}\equiv k^{-1}\{.,.\}
\ee
of the corresponding symplectic form $k\Omega(\p)$,
\be
\{kS_{G/G}(g,A),kS_{G/G}(h,A)\}_{k} = kS_{G/G}(gh,A)-kS_{G/G}(hg,A)\;\;.
\label{gh2}
\ee
Let us now parametrize $g$ as $g=\exp \phi/k$. Then in the $k\ra\infty$
limit the action becomes the BF action $S_{BF}(\phi,A)$,
\be
\lim_{k\ra\infty} kS_{G/G}(g,A) = \trac{1}{2\pi}\int_{\S}\phi F_{A} +
O(k^{-1})\;\;,
\ee
the kinetic term $S_{G}(g,A)$ being of order $O(k^{-2})$ and therefore not
contributing in the limit. Therefore, (\ref{gh2}) becomes
\bea
&&\{S_{BF}(\phi,A),S_{BF}(\phi',A)\} =\nonumber\\
&&\lim_{k\ra\infty} k^{2}(S_{G/G}(\exp\phi/k\exp\phi'/k,A)-
                        S_{G/G}(\exp\phi'/k\exp\phi/k,A))\;\;.\label{gh3}
\eea
Calculating either the left or the right hand side of (\ref{gh3}), one
finds  (\ref{ff}), which is precisely the ordinary equivariance (\ref{equiv2})
of the generator of gauge transformations on $\cA$.

\subs{The Basic Structure of Generalized Hamiltonian Group Actions}

Let us now briefly, and at the risk of being repetitive, extract from the
above the basic structure we have found in the $G/G$ model characterizing
the generalized Hamiltonian group actions and compare it with the standard
theory.

First of all, there is an assignment of vector fields $X(g)$ on a symplectic
manifold $(M,\Omega)$ to elements $g$ of a group $\bH$. These vector fields
satisfy
\be
    [X(g),X(h)] = X(gh)-X(hg)\;\;. \label{gm1}
\ee
This should be thought of as the generalization of the condition
(\ref{xy})
expressing the fact that the vector fields $X_{M}$ provide a realization of
the Lie algebra $\lh$ of $\bH$ on $M$. (\ref{gm1}) can be interpreted as
follows. By linearity one can extend the assignment of vector
fields to elements of $\bH$ to elements of the group algebra $\ZZ\bH$ or
$\CC\bH$ of
$\bH$ so that we can write the right hand side of (\ref{gm1}) as
$X(gh-hg)$. The group algebra can be equipped with a Lie algebra structure by
defining the commutator to be $[g,h] = gh-hg$. Then (\ref{gm1}) can be
read as expressing the fact that the vector fields $X(g)$ provide a
representation of the Lie algebra $(\ZZ\bH,[.,.])$ on $M$,
\be
[X(g),X(h)] = X([g,h])\;\;.\label{gm11}
\ee

Next we demand that these vector fields are Hamiltonian, i.e.~that
there exist functions $F(g)$ on $M$ such that
\be
 i(X(g))\Omega = dF(g)\;\;.     \label{gm2}
\ee
We then write $X(g)=V_{F(g)}$. It follows from (\ref{gm1}) and (\ref{gm2})
that the Hamiltonian vector field corresponding to the Poisson bracket of
two functions $F(g)$ and $F(h)$ is
\be
V_{\{F(g),F(h)\}} = V_{F(gh)}-V_{F(hg)}\;\;,       \label{gm3}
\ee
this being the analogue of (\ref{equiv1}).
We then have a generalized moment map
\be
F:\bH\ra C^{\infty}(M)\;\;,
\ee
which can also be considered as a map
\be
J:M \ra C^{\infty}(\bH)
\ee
via
\be
(J(m))(h) \equiv (F(h))(m)
\ee
(see (\ref{mom1}-\ref{mom3}). Perhaps it will turn out to be
more convenient to regard $J$ as
a map into the distributions on $\bH$. Either way it is natural to say that
the moment map
is equivariant if (\ref{gm11}) or
(\ref{gm3}) can be lifted to hold at the level of Hamiltonian
functions, i.e.~if one has a representation of the Lie algebra
$(\ZZ\bH,[.,.])$ in the Poisson algebra of functions on $M$,
\be
  \{F(g),F(h)\} = F(gh)-F(hg)\equiv F([g,h])\label{gm4}
\ee
(in writing the second equality we have extended functions on $\bH$ to
functions on $\ZZ\bH$ by  linearity).

That this is indeed a reasonable generalization of the ordinary equivariance
condition can be seen by noting that (\ref{gm4}) implies
that the first moments of $F(g)$, defined by
\be
F^{(1)}(X)=\trac{d}{dt}F(\exp tX)|_{t=0}\;\;,
\ee
satisfy the ordinary equivariance condition (\ref{equiv2}),
\be
\{F^{(1)}(X),F^{(1)}(Y)\}=F^{(1)}([X,Y])\;\;.
\ee
This is the counterpart of the $k\ra\infty$ argument we gave in the case
of the $G/G$ action and the above argument could have alternatively been
phrased in similar terms.

The converse, that ordinary equivariance (\ref{equiv2}) implies (\ref{gm4}),
need however not be true as
the generalized equivariance condition implies a whole hierarchy of conditions
on the higher moments
\be
F^{(n)}(X) = (\trac{d}{dt})^{n}F(\exp tX)|_{t=0}
\ee
which are necessary in order for (\ref{equiv2}) to exponentiate to (\ref{gm4}).

There is another way of relating (\ref{gm4}) to ordinary moment maps,
pointed out to us by V.~Fock. Namely, let us associate to a function $F(g)
\in C^{\infty}(M)$ a function $\hat F(\mu,X)\in C^{\infty}(M)$, where
$\mu\in R(\bH)$
labels an irreducible unitary $d_{\mu}$-dimensional representation $V_{\mu}$
of $\bH$ and $X\in{\rm Lie}\bH$, by
\be
\hat F(\mu,X) = -d_{\mu}\int\! dg\, F(g)\Tr_{\mu}(Xg)\;\;.
\ee
It then follows from (\ref{gm4}) and the orthogonality of the traces that
the $\hat F(\mu,X)$ satisfy the Poisson bracket relations
\be
\{\hat F(\mu,X),\hat F(\nu,Y)\} = \d_{\mu\nu}\hat F(\mu,[X,Y])\;\;.
\ee
Thus $\hat F$ can be thought of as an equivariant moment map (in the ordinary
sense) for the direct sum of Lie algebras
\be
\oplus_{\mu\in R(\bH)} \mu({\rm Lie}\bH)\subset
\oplus_{\mu\in R(\bH)} {\rm End\,}V_{\mu}\;\;.
\ee
Modulo analytical problems, $F(g)$ can be recovered from the functions $\hat
F$.

The moment map in the $G/G$ model has a further property, namely its
gauge invariance. This property, however, is linked with a second action
of the group $\bH=\cal G$ on $M=\cA$ (namely via gauge transformations),
and in the general context would take the form $F(g)(m) = F(h^{-1}gh)(h.m)$,
$h.m$ denoting this extra action of $h\in\bH$ on $m\in M$. However, there
seems to be no reason to demand some such property to hold in general,
and we thus take the conditions (\ref{gm1},\ref{gm2}) (and (\ref{gm4}))
to define what we mean by a generalized (equivariant) Hamiltonian group action.

Clearly much remains to be understood about the properties of these
generalized group actions, primarily of course whether this is at
all an interesting structure to consider in general.

\subsubsection*{Acknowledgements}

We would like to thank A.~Alekseev, F.~Delduc,
V.~Fock, L.~Jeffrey and A.~Weinstein for
discussions and helpful remarks, and the people at the
\'Ecole Normale in Lyon and the CPT II in Marseille, where part of this work
was
carried out, for their hospitality.

\appendix

\section{Aspects of Localization in Yang-Mills Theory}

In this section we illustrate the subtleties we encountered in section
3 when adapting the usual localization arguments to the $G/G$ model in the
simpler case of Yang-Mills theory (or, actually, its topological limit, BF
theory).
We will freely make use of the results established in \cite{btdia}
concerning the global issues involved when diagonalizing Lie algebra or group
valued maps without drawing attention to it every time, as these are only of
secondary importance in the issue at stake.

The action we will consider is
\be
S = \int \phi F_{A} - \trac{1}{2}\int \p\p \label{a1}          \;\;,
\ee
which has the equivariant supersymmetry
\be
\d A = \p\;\;,\;\;\;\;\;\;\d\p = d_{A}\phi\label{a2}\;\;.
\ee
This supersymmetry can be used in various ways to localize the theory to
reducible configurations, i.e.\ to solutions $(A_{c},\phi_{c})$ of the
equation
\be
d_{A_{c}}\phi_{c}=0\label{a3}\;\;.
\ee
One way of seeing this is to add to the action a $\delta$-exact term enforcing
this localization in some limit, e.g.
\bea
S^{s} &=& S -s\delta\int\p *d_{A}\phi \nonumber \\
&=& \int (\phi F_{A} - s d_{A}\phi *d_{A}\phi) - \int(\trac{1}{2}
\p\p+s\p*[\p,\phi])\;\;.
\label{a4}
\eea
as $s$ tends to infinity. This is precisely the large $k$ limit of the deformed
$G/G$ action (\ref{ggsact},\ref{ssplit}). If one were to invoke localization
naively, however, one would conclude that the action of the theory reduces to
$\int\phi_{c}F_{A_{c}}$. For fixed $\phi_{c}$, this integral is
independent of $A_{c}$,
\be
d_{A_{c}} \phi_{c} = 0\;\;{\rm and}\;\;d_{A_{c}+X}\phi_{c}=0 \Rightarrow
\int\phi_{c}F_{A_{c}} = \int\phi_{c}F_{A_{c}+X}\;\;, \label{a5}
\ee
so that the $A_{c}$ integral would not be damped and the naive stationary phase
approximation to the path integral diverges. This is the counterpart of the
observation made in section 3 that the gauged WZ term $\Gamma(g_{c},A_{c})$
is independent of $A_{c}$. In order to correctly seperate the
classical from the quantum fields, one needs a convenient parametrization of
the classical fields, i.e.\ the space of solutions to (\ref{a3}). Assuming
that only the main branch of solutions to these equations is relevant, up
to possibly singular gauge transformations the classical solutions can be
parametrized by pairs $(a_{0},\f_{c})$ where $a_{0}$ and $\f_{c}$ are
$\lt$-valued and $\f_{c}$ is constant. Notice that
the condition for $A$ to be a critical point of the vector field
\be
X_{\cA}(\phi) = d_{A}\phi\frac{\d}{\d A}\;\;,
\ee
is also a condition on $\phi$. This is a case not covered directly by the
traditional localization theorems (which tell us nothing about the
$\phi$-integral), and it is then not surprising that a naive application
of localization to the joint $(A,\phi)$ system may lead one astray. Notice
also that there is no other condition
on the torus gauge field $a_{0}$, so that localization does nothing there. This
reflects the fact that localization is empty once one is left with an Abelian
(quadratic) action.

Now a general (generic)
field $\phi$ can always be written in the form $\phi = h\phi^{\lt} h^{-1}$
for some $\lt$-valued field $\phi^{\lt}$. One is thus led to the decomposition
\be
\phi = h\phi^{\lt}h^{-1} = h(\f_{c} + \fb) h^{-1}\label{a6}\;\;,
\ee
where $\fb$ has no constant mode. This is the correct form of the
naive classical-quantum decomposition $\f=\phi_{c}+ \f_{q}$, disentangling
at the same time localization and gauge invariance. Changing variables
from $\f$ to $(h,\f_{c},\fb)$, the bosonic part of the action becomes
\be
S_{BF}^{s}= \int(\f_{c}+\fb)F_{\ah}  -s d_{\ah}(\f_{c}+\fb)*d_{\ah}(\f_{c}+\fb)
\;\;.\label{a7}
\ee
This change of variables also leads to a Jacobian which we write as
\be
\Det[{\rm ad}(\f_{c}+\fb)]|_{\Omega^{0}(\S,\lk)}\;\;,\label{det1}
\ee
the subscripts indicating that this is
a functional determinant on $\lg/\lt$-valued zero-forms.

Note that this action is manifestly gauge invariant under
\be
A\ra A^{g}\;\;,\;\;\;\;\;\;h\ra \g h\label{a8}
\ee
(the latter corresponding to $\f\ra\f^{g}$), as the gauge field $A$ only
appears in the gauge invariant St\"uckelberg combination $\ah$. It is
convenient to split this gauge field into its $\lt$- and $\lg/\lt$-components,
$\ah = \aht + \ahk$, so that one has
\bea
S_{BF}^{s}&=& \int (\f_{c}+\fb)(F_{\aht}+\trac{1}{2}[\ahk,\ahk])\nonumber\\
          &-& s\int d\fb * d\fb + [\ahk,\f_{c}+\fb]*[\ahk,\f_{c} + \fb]
          \label{a9}\;\;.
\eea
At this point the second problem with the naive localization argument is
apparent. Namely, what appears to be the quadratic form for the `quantum field'
$\fb$ can be absorbed into the first term of the action by a shift of $\aht$.
Thus, if one were to scale this quantum field by $\sqrt{s}$ to eliminate
the $s$-dependence from the quadratic term, one would simultaneously kill the
kinetic term for $\fb$ and $\aht$ in the limit $s\ra\infty$ (which is,
as we have seen,
essentially what a straightforward implementation of localization would lead
one to believe).

The crux of the matter is of course that, as argued above,
localization applies {\em a priori} only to the gauge field integral.
But as this localization is $\phi$ dependent, one needs a
good parametrization of the $\phi$'s to implement this localization
correctly. Arguments based on $s$-independence alone
are nevertheless fine as long as one
makes sure that one keeps quadratic forms for all the fields involved. It is
precisely to
ensure this and to avoid pitfalls like the above
that it is helpful to use an explicit parametrization of the (gauge orbits of)
classical configurations.

We now split $\aht$ into a (possibly non-trivial) background gauge field
$A_{0}$ such that the components of $dA_{0}$ are harmonic,
and a $\lt$-valued one-form $a^{\lt}$ and shift $a^{\lt}$ by $s*d\fb$.
This decouples $(A_{0},\f_{c})$ from $(a^{\lt},\fb)$ and the sole effect
of integrating over $a^{\lt}$ is now to set $\fb$ to a constant and hence to
zero as, by assumption, $\fb$ has no constant mode. Reintroducing the
fermionic fields, one is thus left with the action
\bea
S_{eff}^{s}&=& 2\pi \f_{c}.n + \int(\trac{1}{2}\f_{c}[\ahk,\ahk]
-s[\ahk,\f_{c}]*[\ahk,\f_{c}])\nonumber\\
&-&\int (\trac{1}{2}\p^{h}\p^{h} +  s \p^{h}*[\p^{h},\f_{c}])\;\;,\label{a10}
\eea
the first term representing the pairing between the constant field $\f_{c}$
and the $r$-tuple of integers characterizing the first Chern class $[dA_{0}]$
of $\aht$ in $H^{2}(\S,\ZZ^{r})\sim \ZZ^{r}$.

Note again that this action is still gauge invariant and that no localization
or approximation has entered into the derivation of (\ref{a10}).
One can now proceed in a number of ways to evaluate the partition
function, each one of them also being more or less readily available in the
$G/G$ model. It is here that one has the choice between solving the
theory by Abelianization or by localization, but the following discussion
should make it clear that at this point the distinction between the two methods
is rather artificial. This illustrates once more the main point we wanted to
make in section 3 in the context of the $G/G$ model, namely that localization
abelianizes the theory (the converse having already been established in
\cite{btver,btlec}).

\begin{enumerate}
\item As everything is still independent of $s$ and well defined for $s=0$,
one can simply set $s$ equal to zero. One is then just left with the original
theory, expressed in terms of $\f_{c}$ and $\ahk$, $\aht$ and $\fb$ having
been integrated out. The group valued field $h$ just represents the gauge
degrees of freedom and has to be dealt with in some way:
\begin{enumerate}
\item Performing the change of variables $A\ra \ah$, the $h$-integral
becomes the gauge volume and factors out. The integral over
$A^{\lg/\lt}$ produces
the functional determinant
\be
\Det^{-1/2}[{\rm ad}\f_{c}]|_{\Omega^{1}(\S,\lk)}\;\;.\label{det2}
\ee
Combined with the Jacobian (\ref{det1}) from the change of variables,
this gives the residual finite-dimensional determinant
\be
\det{}^{\c(\S)/2}[{\rm ad}\f_{c}]|_{\lk}\;\;,\label{det3}
\ee
($\c(\S)$ denoting the Euler number of $\S$)
leading to the standard result for the partition function of Yang-Mills theory
upon summation over all topological sectors and performing the
finite-dimensional integral over $\f_{c}$ \cite{btlec}.
\item One can also choose the gauge $h=1$ (this is Abelianization).
This obviously has the same effect as the above change of variables.
\item Lastly, one can of course choose any other gauge condition as well, e.g.\
a covariant gauge, and still do all the integrals explicitly. The integrals
over $h$, the ghosts and the Lagrange multiplier enforcing the gauge condition
combine to give 1 (by running the Faddeev-Popov trick backwards), reducing one
to possibility (a).
\end{enumerate}
\item
Alternatively, one can consider the limit $s\ra\infty$ (localization). To that
end one scales the quantum fields $\ahk$ and their superpartners
$(\p^{h})^{\lg/\lt}$ as
\be
\ahk\ra s^{-1/2}\ahk\;\;,\;\;\;\;\;\;(\p^{h})^{\lg/\lt}\ra
s^{-1/2}(\p^{h})^{\lg/\lt}
\;\;.\label{a11}
\ee
In the limit $s\ra\infty$, the terms coming form the original BF theory and
involving $\ahk$ disappear and one obtains
\bea
S_{eff}^{s\ra\infty}&=&2\pi\f_{c}.n -\int_{\S} [\ahk,\f_{c}]*[\ahk,\f_{c}]
\nonumber\\
&-&\int_{\S} (\p^{h})^{\lg/\lt}*[(\p^{h})^{\lg/\lt},\f_{c}] +\trac{1}{2}
\p^{\lt}\p^{\lt}\;\;.\label{a12}
\eea
Again the $h$-integral can be dealt with in several ways. For simplicity
we will follow option 1(a) above and perform the change of variables $A\ra\ah$.
Then one finds that the integrals over $A^{\lg/\lt}$ and $\p^{\lg/\lt}$ give
\be
\Det^{-1/2}[({\rm ad}\f_{c})^{2}]_{\Omega^{1}(\S,\lk)} \label{det4}
\ee
and
\be
\Det^{1/2}[{\rm ad}\f_{c}]|_{\Omega^{1}(\S,\lk)} \label{det5}
\ee
respectively, combining to give the net contribution (\ref{det2}), in
agreement with the result obtained in 1(a).
\item In this example it is also straightforward to work out what happens
for finite values of $s$. Once again with $A\ra\ah$ for simplicity,
one finds that the quadratic terms in $A^{\lg/\lt}$ and $\p^{\lg/\lt}$ are of
the form
\[A^{\lg/\lt}{\rm ad}\f_{c} (1-2s\; {\rm ad}\f_{c} *) A^{\lg/\lt} +
 \p^{\lg/\lt}(1-2s\; {\rm ad}\f_{c} *)\p^{\lg/\lt}\;\;.\]
Evidently this also leads to the same net determinant (\ref{det2}),
establishing explicitly the $s$-independence of the theory.
\end{enumerate}

\rnc{\Large}{\normalsize}


\begin{thebibliography}{00}
\addcontentsline{toc}{section}{References}
\frenchspacing
\small
\addtolength{\itemsep}{-4pt}
\bibitem{btver} M. Blau and G. Thompson, {\em Derivation of the Verlinde
                Formula from Chern-Simons Theory and the $G/G$ model},
                Nucl. Phys. B408 (1993) 345-390.
\bibitem{ver} E. Verlinde, {\em Fusion rules and modular transformations in
            $2d$ conformal field theory}, Nucl. Phys. B300 (1988) 360-376.
\bibitem{ger} A. Gerasimov, {\em Localization in GWZW and Verlinde Formula},
               Uppsala preprint UUITP-16/1993, 15 p., hep-th/9305090.
\bibitem{btlec} M. Blau and G. Thompson, {\em Lectures on 2d Gauge Theories:
                Topological Aspects and Path Integral Techniques}, ICTP
                preprint IC/93/356, 70 p.,  hep-th/9310144.
                To appear in the Proceedings of the 1993 Trieste Summer
                School on High Energy Physics and Cosmology.
\bibitem{ew2d} E. Witten, {\em Two dimensional gauge theories revisited}, J.
               Geom. Phys. 9 (1992) 303-368.
\bibitem{verver} E. Verlinde and H. Verlinde, {\em Conformal field theory and
                geometric quantization}, in {\em Superstrings '$89$} (eds. M.
                Green et al.), World Scientific, Singapore (1990) 422-454.
\bibitem{btdia} M. Blau and G. Thompson, {\em On diagonalization in ${\rm Map}
                (M,G)$}, ICTP preprint IC/94/37, 41 p.,
                hep-th/9402097.
\bibitem{ew}    E. Witten, {\em The Verlinde Algebra and the Cohomology of
                the Grassmannian}, IAS preprint IASSNS-HEP-93/41, 78 p.,
                 hep-th/9312104.
\bibitem{bismut} J.-M. Bismut, {\em Equivariant Bott-Chern currents
      and the Ray-Singer analytic torsion}, Math. Ann. 287 (1990) 495-507.
\bibitem{beau}  A. Beauville, {\em Vector bundles on curves and generalized
                theta functions: recent results and open problems}, preprint
                15 p.,  alg-geom/9404001; and
                {\em Conformal blocks,
                fusion rules and the Verlinde formula}, preprint 24 p.,
                 alg-geom/9405001.
\bibitem{szenes} A. Szenes, {\em The combinatorics of the Verlinde formula},
                preprint 13 p.,  alg-geom/9402003.
\bibitem{jk} L.C. Jeffrey and F.C. Kirwan, {\em Localization for non-Abelian
     group actions}, Oxford preprint 36 p.,  alg-geom/9307001;
     and {\em in preparation}.
\bibitem{ewwzw} E. Witten, {\em On holomorphic factorization of WZW and coset
                models}, Commun. Math. Phys. 144 (1992) 189-212.
\bibitem{bv} N. Berline and M. Vergne, {\em Classes charact\'eristiques
             \'equivariantes. Formule de localisation en cohomologie
             \'equivariante}, C. R. Acad. Sci. Paris 295 (1982) 539-541;
             N. Berline, E. Getzler and M. Vergne, {\em Heat Kernels and Dirac
             Operators}, Springer Verlag (1992).
\bibitem{dh} J.J. Duistermaat and G. Heckman, {\em On the variation in the
             cohomology of the symplectic form of the reduced phase space},
             Inv. Math. 69 (1982) 259-268, ibid. 72 (1983) 153-158.
\bibitem{bgs} J.-M. Bismut, H. Gillet and C. Soul\'e, {\em Analytic torsion
             and holomorphic determinant bundles I. Bott-Chern forms and
             analytic torsion}, Commun. Math. Phys. 115 (1988) 49-78.
\bibitem{vag} V.A. Ginzburg, {\em Equivariant cohomology and K\"ahler's
            geometry}, Funct. Anal. App. 21 (1987) 271-283.
\bibitem{bc} R. Bott and S. Chern, {\em Hermitian vector bundles and the
             equidistribution of the zeros of their holomorphic sections},
             Acta Math. 114 (1968) 71-112.
\bibitem{jaap} J. Kalkman, {\em BRST model for equivariant cohomology and
              representatives for the equivariant Thom class}, Commun. Math.
              Phys. 153 (1993) 447-463.
\end{thebibliography}
\end{document}